\documentclass[aps,nofootinbib,amsmath,reprint,floatfix,superscriptaddress,twocolumn]{revtex4}
\usepackage{amssymb,color,bbm,mathrsfs,esvect,amsmath,graphicx,latexsym,amsthm,slashed,eso-pic,amsfonts,bbold}
\usepackage[dvipsnames]{xcolor}
\definecolor{darkgreen}{rgb}{0,0.6,0}
\definecolor{darkblue}{rgb}{0,0,0.5}
\usepackage[colorlinks=true,linkcolor=black,citecolor=black,urlcolor=darkblue]{hyperref}
\usepackage[final]{pdfpages}
\newcommand{\be}{\begin{equation}}
\newcommand{\ee}{\end{equation}}
\newcommand{\nl}{\nonumber \\}

\usepackage{verbatim}
\usepackage{ulem}

\newcommand{\GeV}{\text{ GeV}}
\newcommand{\MeV}{\mathrm{ MeV}}
\newcommand{\keV}{\mathrm{ keV}}

\newcommand{\Ap}{{A^\prime}}
\newcommand{\mAp}{m_{A^\prime}}
\newcommand{\order}[1]{\mathcal{O}(#1)}
\newcommand{\lag}{\mathcal{L}}
\newcommand{\Tfo}{T_{\mathrm{fo}}}
\newcommand{\Ebeam}{E_{\mathrm{beam}}}

\newcommand{\SLAC}{SLAC National Accelerator Laboratory, 2575 Sand Hill Road, Menlo Park, CA, 94025, USA}
\newcommand{\Cincinnati}{Department of Physics, University of Cincinnati, Cincinnati, Ohio 45221, USA}
\DeclareMathOperator{\tr}{Tr}

\def\lsim{\mathrel{\raise.3ex\hbox{$<$\kern-.75em\lower1ex\hbox{$\sim$}}}}
\def\gsim{\mathrel{\raise.3ex\hbox{$>$\kern-.75em\lower1ex\hbox{$\sim$}}}}

\begin{document}

\hspace{13cm} \parbox{5cm}{SLAC-PUB-17135}~\\

\hspace{13cm}

\title{Cosmology and Accelerator Tests of Strongly Interacting Dark Matter}
\author{Asher Berlin} 
\author{Nikita Blinov}
\affiliation{\SLAC}
\author{Stefania Gori}
\affiliation{\Cincinnati}
\author{Philip Schuster}
\author{Natalia Toro} 
\affiliation{\SLAC}

\date{\today}

\begin{abstract}
  A natural possibility for dark matter is that it is composed of the stable pions of a QCD-like hidden sector. Existing literature largely assumes that pion self-interactions alone control the early universe cosmology. We point out that processes involving vector mesons typically dominate the physics of dark matter freeze-out and significantly widen the viable mass range for these models. The vector mesons also give rise to striking signals at accelerators. For example, in most of the cosmologically favored parameter space, the vector mesons are naturally long-lived and produce Standard Model particles in their decays. Electron and proton beam fixed-target experiments such as HPS, SeaQuest, and LDMX can exploit these signals to explore much of the viable parameter space. We also comment on dark matter decay inherent in a large class of previously considered models and explain how to ensure dark matter stability.
\end{abstract}

\maketitle

\section{Introduction\label{sec:intro}}

Despite the wealth of gravitational evidence for dark matter (DM), its nature 
remains a fundamental puzzle in particle physics. 
If DM possesses non-negligible interactions with the
inflaton or Standard Model (SM) bath, a large thermal population is necessarily generated 
in the early universe. This abundance must be depleted to match the observed DM energy density. In the Weakly Interacting
Massive Particle (WIMP) paradigm, this occurs through annihilations
to SM particles. In general, however, once the temperature drops below the
DM mass, any process that reduces the DM number density can potentially lead to
a viable freeze-out scenario. 

A simple and generic possibility is that DM is the lightest state of a strongly coupled hidden sector (HS) that is uncharged under SM forces. 
In this case, the dynamics of the HS opens up qualitatively new mechanisms for depleting the DM abundance. The best-known possibility is where annihilations of three DM particles into two DM particles deplete the thermal abundance~\cite{Carlson:1992fn}, as realized in models of Strongly Interacting Massive Particles (SIMPs)~\cite{Hochberg:2014dra,Hochberg:2014kqa}. Nonetheless, to be cosmologically viable, the HS must couple to the SM in order to maintain kinetic equilibrium, though the required coupling is significantly weaker than would be needed to achieve the DM abundance through pair-annihilation.

We focus on a class of strongly interacting hidden sectors with QCD-like particle content.
Dark matter is
composed of the lightest states in this sector, $\pi_D$ -- the pseudo-Nambu-Goldstone
bosons of a spontaneously broken chiral symmetry, similar to the pions of
QCD~\cite{Hochberg:2014kqa,Bai:2010qg,Belyaev:2010kp,Buckley:2012ky,Cline:2013zca,Antipin:2015xia,Davoudiasl:2017zws}. Kinetic equilibrium between the dark sector 
and the SM can be maintained through interactions via a light 
mediator, such as a dark photon, $\Ap$, the gauge boson of a 
broken $U(1)_D$ symmetry that kinetically mixes with SM hypercharge.
The strongly interacting sector necessarily includes heavier 
excitations, such as vector mesons, $V_D$, the analogues of SM $\rho$ mesons, which are expected to mix with the dark photon.

These vector mesons, $V_D$, have important consequences for both the cosmology and phenomenology of SIMPs.
Past studies have largely ignored the impact of vector mesons on SIMP cosmology. This is justified in the chiral limit, where $V_D$'s are parametrically heavier than DM pions.  However, SIMP models rely on significant chiral symmetry breaking to achieve sufficiently large $3\pi_D \rightarrow 2\pi_D$ cross-sections, so that one expects $m_{V_D} \sim m_{\pi_D}$.  In this case, $V_D\rightarrow \pi_D\pi_D$ decays are kinematically forbidden, and the semi-annihilation~\cite{DEramo:2010keq} reaction, $\pi_D \pi_D \rightarrow \pi_D V_D$ followed by $V_D \to \text{SM}$, is often the primary process responsible for depleting the DM abundance.   Accounting for this process consistently dramatically expands the viable range for the DM mass and 
the coupling to the SM. We find that pion masses ranging from $\sim 10 \ \MeV$ to well above the GeV-scale are allowed, in contrast to the narrow mass window around $\sim 100 \ \MeV$ when only $3\rightarrow 2$ reactions are considered~\cite{Hochberg:2014dra,Hochberg:2014kqa}. 
This process also allows the coupling between the HS and SM to be orders of magnitude smaller than previously thought.

\begin{figure*}[t]
\centering
\hspace{-0.5cm}
\includegraphics[width=15.5cm]{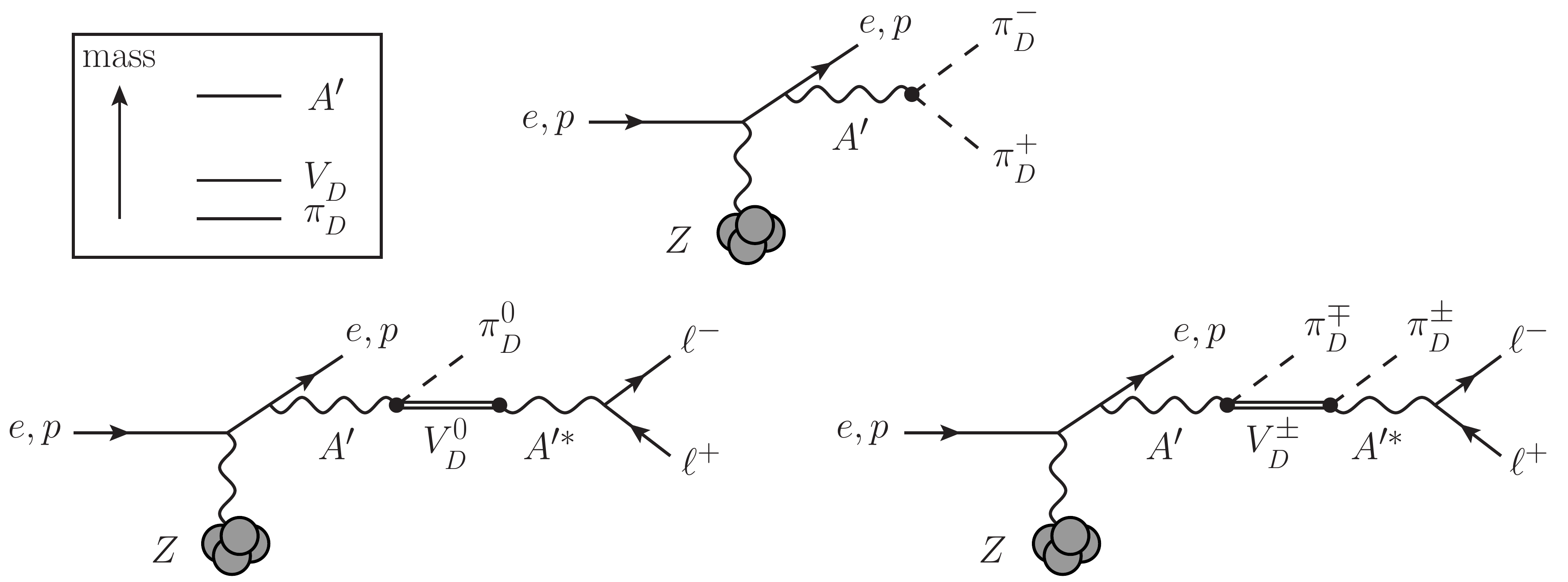} \hspace{-0.5cm}
\caption{Hidden sector particle production and decay at electron and proton fixed-target experiments. Dark photons ($\Ap$) are produced via Bremsstrahlung in an electron- or proton-nucleus collision and decay promptly into a pair of hidden sector pions ($\pi_D$), as shown in the top diagram, or a pion and vector meson ($V_D$) of the hidden sector, as shown in the bottom diagrams. At proton beam experiments, dark photons are also produced through Standard Model meson decays and Drell-Yan (not shown).
The vector meson is long-lived, decaying into Standard Model leptons (through mixing
with the $\Ap$) after traversing a macroscopic distance from the target (bottom-left). 
Similar processes can also occur for non-singlet vector
mesons which undergo a three-body decay (through an off-shell $\Ap$) into a hidden sector pion and a pair of Standard Model leptons (bottom-right).
The inset shows a schematic hidden sector mass spectrum, with $\mAp / 2 \gtrsim
m_{V_D} \sim m_{\pi_D}$, which enables these decays.
} \label{fig:feyndiag}
\end{figure*}

A visibly decaying vector meson also opens up significant new avenues for accelerator-based dark sector searches.  Some of these have been discussed in Refs.~\cite{Essig:2009nc,Hochberg:2015vrg,Hochberg:2017khi}. We focus here on the visible signals associated with dark photon decays to the strongly-coupled sector.  A striking signature of a strongly interacting HS is the decay of a dark photon $\Ap \rightarrow \pi_D V_D$ with subsequent resonant decay $V_D\rightarrow \ell^+\ell^-$ (where $\ell$ is a SM lepton), as shown in the bottom-left diagram of Fig.~\ref{fig:feyndiag}. 
For natural parameter values, there is an observably displaced $V_D$-decay vertex. This signal differs from the direct decay of a dark photon (or direct production of the $V_D$) in the energy spectrum of the decay products (see Fig.~\ref{fig:kin}). More dramatically, the production rate for $V_D$'s is parametrically larger than that of an $\Ap$ for a given lifetime.  A related process is the decay of the dark photon into vector mesons whose quantum numbers do not permit mixing with the dark photon, as shown in the bottom-right diagram of Fig.~\ref{fig:feyndiag}. 
These vector mesons decay to $\pi_D \ell^+ \ell^-$ final states with even longer lifetimes.

These distinctive signatures can be searched for at beam dump and fixed-target experiments. Such searches are complementary to the minimal signals of HS DM, e.g., nuclear/electron recoils and invisible dark photon decays, the latter of which is shown in the top diagram of Fig.~\ref{fig:feyndiag}. Data from the E137 beam dump experiment is already able to probe interesting regions of parameter space, especially for $\sim 100 \text{ meter}$ decay lengths. 
Complementary viable regions will be tested in the near future at the currently running Heavy Photon Search (HPS) experiment, an upgrade of the SeaQuest experiment,
and at the proposed Light Dark Matter eXperiment (LDMX).
Our main 
results are summarized in Fig.~\ref{fig:EpsAndFpi}, where
we show existing constraints as well as sensitivity of HPS, SeaQuest, and LDMX to  
cosmologically-motivated models that have not been tested otherwise. 
Similar signals are also observable above the muon threshold 
at the $B$-factories BaBar and Belle-II and at the Large Hadron Collider (LHC).

This paper is organized as follows. In Sec.~\ref{sec:model}, we describe a benchmark model 
of a strongly interacting HS that we use throughout this work.
We also show that HS vector mesons are long-lived for well-motivated parameter values and, 
therefore, can give rise to displaced vertex signals at fixed-target and collider experiments.
In Sec.~\ref{sec:cosmo}, we discuss the cosmological importance of these vector mesons 
and clarify the issue of pion stability. We then demonstrate in Sec.~\ref{sec:signals} that 
existing and future fixed-target, collider, and direct detection experiments are sensitive to 
cosmologically-motivated parameter space. We also briefly comment on
various astrophysical and cosmological probes.
Finally, we summarize our conclusions in Sec.~\ref{sec:conclusion}. Details of the model, cross-sections and decay rates, and Boltzmann 
equations are provided in Appendices~\ref{app:model}--\ref{app:cosmo}.

\section{A Strongly Interacting Sector\label{sec:model}}

We consider a strongly interacting HS described by a confining $SU(N_c)$ gauge
theory with $N_c=3$ colors, analogous to SM QCD. We also introduce $N_f$ light
flavors of Dirac fermions in the fundamental representation. We are interested
in the relative importance of $3 \pi_D \to 2 \pi_D$ and $\pi_D
\pi_D \to \pi_D V_D$ in dictating the DM abundance. We choose $N_f = 3$, as
this is the minimum number of flavors that is required to allow either process. 
In this section, we briefly outline the basics of the model, while a more
detailed discussion is provided in Appendix~\ref{app:model}. Hereafter, we
denote the HS pions and vector mesons as $\pi$ and $V$, respectively (a
subscript ``$D$" is implied). For $\pi$ and $V$, the superscripts, $0$ and
$\pm$, denote charges under $U(1)_D$, while for $\ell$, they denote charges
under $U(1)_\text{em}$. The global chiral symmetry, $SU(N_f)_L \times
SU(N_f)_R$, is spontaneously 
broken by the hidden quark condensate to the diagonal subgroup, $SU(N_f)_V$, during confinement.
Thus, at low energies this is a 
theory of $N_f^2-1$ pions, $\pi$, which constitute the DM of the universe. 
The low-energy pion self-interactions are described by chiral perturbation
theory; the strength of these interactions is characterized by the HS pion decay constant, $f_\pi$.\footnote{Note that our convention for $f_\pi$ differs from that of Refs.~\cite{Hochberg:2014kqa,Hochberg:2015vrg} 
by a factor of two.
}
The odd intrinsic-parity
Wess-Zumino-Witten term of the chiral Lagrangian
enables $3 \to 2$ pion annihilations~\cite{Wess:1971yu,Witten:1983tw}. If such
processes are the dominant pion number-changing mechanism during thermal
freeze-out, the HS pion abundance matches the observed DM energy density for
$m_\pi / f_\pi \gtrsim \text{few}$~\cite{Hochberg:2014kqa}. On the other hand,
observations of merging galaxy clusters lead to upper limits on the DM
self-scattering cross-section, at the level of $\sigma_\text{scatter} / m_\pi
\lesssim \text{few} \times \text{cm}^2 /
\text{g}$~\cite{Clowe:2003tk,Markevitch:2003at,Randall:2007ph}. As shown in previous studies, these 
considerations suggest that $m_\pi \sim \text{few} \times 100
\ \MeV$ and $m_\pi / f_\pi \sim 4 \pi$~\cite{Hochberg:2014kqa}.

Vector mesons, $V$, are expected to appear in the spectrum at a scale
close to the cutoff of the theory, $m_V \sim 4 \pi \, f_\pi/\sqrt{N_c}$~\cite{Soldate:1989fh,Chivukula:1992gi,Georgi:1992dw}. 
Hence, for large values of $m_\pi/ f_\pi$, the DM pions are
parametrically close in mass to the lightest spin-1 resonances of the HS. In
particular, for the $3\rightarrow 2$ cosmologically-motivated values, $m_\pi/ f_\pi\gtrsim 4$, 
the vector mesons are expected to lie near the two pion threshold, $m_V \sim 2 \, m_\pi$. 
If $m_V < 2 \, m_\pi$, the decays $V\rightarrow \pi\pi$ are kinematically forbidden and instead $V$ must 
decay to SM states. As a result, vector mesons play an important role in the cosmological depletion of DM pions. 

Considerations of structure formation in the early universe demand that kinetic equilibrium between the HS and the SM bath must be
maintained during freeze-out~\cite{Hochberg:2014dra}.  A light mediator that couples weakly to the SM
can allow for efficient heat transport between the two sectors.  The dark
photon, $\Ap$, is an ideal candidate for this task since it interacts with the SM
through a renormalizable operator; thus, if its interactions are not too feeble, it is able to maintain kinetic
equilibrium between the two sectors even at low freeze-out temperatures relevant
for light SIMPs (see also Ref.~\cite{Evans:2017kti} for a recent investigation of HS-SM
equilibration). The $\Ap$ is a gauge boson of a broken $U(1)_D$
symmetry that couples to the SM through kinetic mixing with the hypercharge 
gauge boson, $B$,
\be
\label{eq:kinmix}
\lag \supset \frac{\epsilon}{2 \cos{\theta_W}} ~ A_{\mu \nu}^\prime \, B^{\mu \nu}
~,
\ee
where $A_{\mu \nu}^\prime$ and $B_{\mu \nu}$ are the $U(1)_D$ and $U(1)_Y$ field strengths, respectively, and $\theta_W$ is the Weinberg angle~\cite{Okun:1982xi,Holdom:1985ag}. 
The kinetic mixing parameter, $\epsilon$, can be generated 
by loops of heavy particles charged under both $U(1)_D$ and $U(1)_Y$. 
Its natural size is therefore $\lesssim 10^{-2}$ ($\lesssim 10^{-4}$) if 
it is generated at one (two) loop(s)~\cite{Holdom:1985ag,delAguila:1988jz}.
For $\mAp \ll 100\GeV$, the $\Ap$ dominantly mixes with the SM photon. After electroweak 
symmetry breaking, SM fermions acquire a millicharge under $U(1)_D$, given
by $\epsilon \, e \, Q_f$, where $e$ is the electromagnetic coupling and $Q_f$
is the electric charge of the SM fermion in units of $e$. Dark photon 
masses below the electroweak scale naturally arise in, e.g., supersymmetric 
theories in which the HS is sequestered from the source 
of supersymmetry breaking~\cite{Morrissey:2009ur}. In addition, $\mAp \sim \epsilon \times \order{100} \GeV$ is independently motivated in theoretical constructions where electroweak symmetry breaking triggers $U(1)_D$ breaking at the GeV-scale~\cite{Cheung:2009qd}. With $\Ap$ at the GeV-scale, different mass hierarchies for the other HS states are possible. From a bottom-up perspective, $\mAp \gtrsim m_\pi$ is required to prevent observable distortions of the cosmic microwave background (CMB) through late-time DM annihilations of the form $\pi \pi \to \Ap \Ap$ followed by $\Ap \to 2 \ell$~\cite{Ade:2015xua}. Hence, we will focus on $m_\pi \sim m_V \lesssim  \mAp 
 \sim \text{GeV}$. 

Interactions between the $\Ap$ and HS mesons are incorporated by
embedding $U(1)_D$ in the unbroken $SU(N_f)_V$ subgroup of the global chiral 
symmetry. This embedding is specified 
by the HS quark $U(1)_D$ charge assignments, defined by 
the charge matrix, $Q$. 
A generic choice of $Q$ enables decays of neutral HS 
pions into SM leptons via tree-level ($\pi \to A^{\prime *} A^{\prime *} \to 4 \ell$) and loop-induced ($\pi \to 2 \ell$) processes.
The leading contribution to these processes in chiral perturbation theory 
is controlled by the chiral anomaly, which is proportional to $\tr (T_\pi \, Q^2)$, where
$T_\pi$ is the $SU(N_f)$ generator corresponding to $\pi$.
If the neutral pions are unstable during DM freeze-out at $t\sim H(\Tfo)^{-1}$, 
charged pions can scatter into these states, 
efficiently depleting the entire DM abundance to a negligible value~\cite{Hochberg:2015vrg}. 
Therefore, the viability of this DM scenario requires 
the neutral pions to be long-lived on timescales compared to freeze-out.
The leading contribution to neutral pion decay vanishes if $Q^2 \propto \mathbb{1}$ since $\tr T_\pi = 0$. 
Following Refs.~\cite{Lee:2015gsa,Hochberg:2015vrg}, we take 
\be
\label{eq:Q}
Q = \text{diag}(+1, -1, -1)
~.
\ee
It has been claimed that this charge assignment is sufficient to make 
all DM pions stable~\cite{Lee:2015gsa,Hochberg:2015vrg}. However, the anomalous term is only the leading contribution 
to the decay; higher dimensional operators in chiral perturbation theory 
also contribute to this process~\cite{Bijnens:1988kx}.\footnote{We thank Markus Luty and Jack Gunion for emphasizing this point to us.}
Indeed, even in the absence of the
electromagnetic chiral anomaly, the SM $\pi^0_\mathrm{SM}$ would be unstable but with a much longer lifetime~\cite{Donoghue:1992dd}. 
Therefore, for general quark mass assignments, only the charged HS pions are stable since they are the 
lightest states protected by the global $U(1)_D$ symmetry.
In Sec.~\ref{sec:cosmo}, we argue that these states can be viable DM 
candidates even if the neutral pions are unstable.

Hidden sector mesons can be created in terrestrial experiments through their interactions with the $\Ap$.
Gauge symmetry determines the interactions of $\Ap$ with the charged 
mesons, while the neutral vector mesons mix with the $\Ap$ through interactions of the form~\cite{Klingl:1996by,Harada:2003jx,Hochberg:2015vrg}
\be
\mathcal{L} \supset - \, \frac{e_D}{g} ~ A_{\mu \nu}^\prime ~ \tr \, Q \, V_{\mu \nu}
~,
\label{eq:vector_meson_mixing}
\ee
where $V_{\mu\nu}$ is the vector meson field strength, $e_D$ is the $U(1)_D$ gauge coupling, and $g$ is the 
$V\pi\pi$ interaction strength. We relate $g$ to the vector meson mass and pion decay constant using the KSRF relation\footnote{In the Hidden Local Symmetry framework of vector meson-pion interactions, the KSRF relation follows from the Vector Meson Dominance limit~\cite{Bando:1984ej}, where the $\Ap$ couples to $\pi$ only through mixing with an intermediate $V$.}~\cite{Kawarabayashi:1966kd,Riazuddin:1966sw},
\be
g = \frac{m_V}{\sqrt{2} \, f_\pi}
~.
\ee
Anomalous decays, such as $\Ap \rightarrow \pi V$, are described by the gauged Wess-Zumino-Witten interactions~\cite{Kaymakcalan:1983qq}.  

\begin{figure}[t]
  \centering
\includegraphics[width=8.5cm]{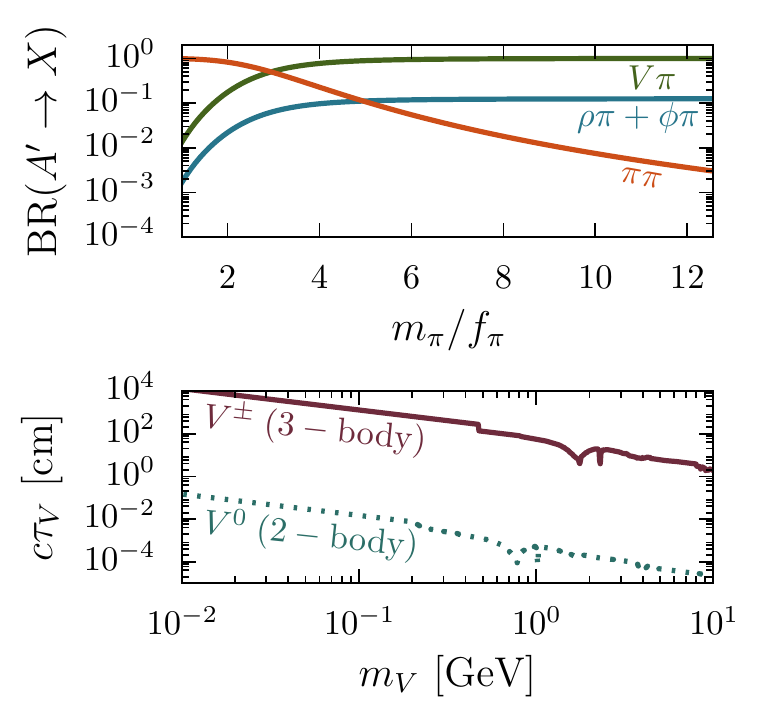} 
\caption{
(Upper panel) Branching fraction of the $\Ap$ into hidden sector mesons as a function of the
ratio $m_\pi / f_\pi$ for $\mAp / m_\pi = 3$, $m_V/m_\pi = 1.8$, and $\epsilon \ll 1$. (Lower panel) The proper lifetime of 
neutral (dotted line) and charged (solid line) vector mesons for $\alpha_D =10^{-2}$, $\epsilon = 10^{-3}$, and $m_\pi / f_\pi = 3$.
} \label{fig:ApBF}
\end{figure}

An on-shell $\Ap$ can be radiated in an electron- or proton-nucleus collision as depicted
in Fig.~\ref{fig:feyndiag}. It can also be produced through SM meson decays and Drell-Yan at proton beam fixed-target experiments.
 The observable signals are determined by its decays.
For $\mAp > 2 \,m_\pi$ and $\epsilon \ll 1$, the $\Ap$ 
decays almost exclusively into HS mesons,
\begin{align}
& \Gamma (\Ap \to \pi \pi) \propto \alpha_D  ~ \mAp
\nl
& \Gamma (\Ap \to \pi V) \propto \alpha_D ~ g^2 ~ \mAp^3 / f_\pi^2
~,
\end{align}
where $\alpha_D \equiv e_D^2/ 4\pi$. 
Since the choice of $Q$ in Eq.~\eqref{eq:Q} is structurally similar to that of the SM, we adopt the standard meson naming scheme (see also Appendix \ref{app:model}).\footnote{Note, however, that the HS pions have $U(1)_D$ charges $0 , \pm 2$ for the choice in Eq.~\eqref{eq:Q}.}
We present the branching ratios of the $\Ap$
to various HS final states in the upper panel of Fig.~\ref{fig:ApBF}, as a function of
$m_\pi / f_\pi$ for $\mAp / m_\pi = 3$, $m_V/m_\pi = 1.8$, and $\epsilon\ll 1$.
For $m_\pi / f_\pi \gtrsim 3$, the anomalous decays, $\Ap \to V \pi$, dominate the total $\Ap$ width, while for
$m_\pi / f_\pi \lesssim 3$, the $\Ap$ has a significant invisible branching fraction into 
pairs of charged pions (and charged vector mesons if $m_\Ap > 2 \, m_V$). 

The vector mesons are 
unstable. If $V$ mixes with the $\Ap$, i.e., $\tr(Q \, T_V) \neq 0$ 
(see Eq.~\eqref{eq:vector_meson_mixing}), and $m_V < 2 \, m_\pi$, $V$ decays directly to the SM. 
The decay rate into a pair of SM leptons is of the form
\be
\label{eq:Vdecay2body}
\Gamma(V \rightarrow \ell^+ \ell^-) \propto \frac{\alpha_D ~ \alpha_\text{em} ~ \epsilon^2}{g^2} ~
\frac{m_V^5}{\mAp^4} 
~.
\ee
The corresponding decay length can easily be macroscopic 
since $\Gamma$ is suppressed by the small parameters $\epsilon$ and $\alpha_D/g^2$. 
The spin-1 $U(1)_D$-neutral states $\rho$ and $\phi$ decay to the SM in this manner, 
while $\omega - \Ap$ mixing vanishes at leading order.
As a result, the $\omega$ and charged vector mesons decay
via three-body processes through an off-shell $\Ap$,  
e.g., $\omega \to \pi \, A^{\prime *} \to \pi \, \ell^+\ell^-$ 
and $V^\pm \to \pi^\pm A^{\prime *} \to \pi^\pm \ell^+ \ell^-$. 
In the limit that $m_\ell \ll m_\pi \ll m_V \ll \mAp$, these decay rates take the parametric form
\be
\Gamma(V^\pm\rightarrow \pi^\pm \ell^+ \ell^-) \propto \alpha_D ~ \alpha_\text{em} ~ \epsilon^2 ~ g^2 ~ \frac{m_V^7}{f_\pi^2 ~ \mAp^4}
~,
\ee
and are further suppressed by three-body phase space factors.
The corresponding decay lengths are typically much larger 
than those of the two-body decays in Eq.~\eqref{eq:Vdecay2body}.\footnote{Analogous three-body decays to SM states and DM also occur in supersymmetric hidden 
sectors and provide a powerful probe of these models 
at fixed-target experiments~\cite{Morrissey:2014yma}.}
The decay lengths for two- and 
three-body decays are shown in the lower panel of Fig.~\ref{fig:ApBF} for 
$\alpha_D =10^{-2}$, $\epsilon = 10^{-3}$, and $m_\pi / f_\pi = 3$ as a function of the vector 
meson mass. We note that, even in the two-body case, $V$ has a macroscopic 
decay length for $m_V \lesssim \text{GeV}$. Full expressions for the decays considered above are given in Appendix~\ref{app:dec}.

As shown in the top panel of Fig.~\ref{fig:ApBF}, HS vector mesons are produced with an $\mathcal{O}(1)$ 
branching fraction in $\Ap$ decays for $m_\pi / f_\pi \gtrsim \text{few}$. These vector mesons are parametrically 
long-lived compared to the $\Ap$ if $m_V < 2 \, m_\pi$. Thus, if such vectors are produced, their decays to the SM are naturally
displaced, giving rise to unique experimental signatures.
Higher-order corrections to meson masses from chiral symmetry breaking effects 
can give rise to various spectra where only a subset of the 
vector meson masses lie below $2 \, m_\pi$. 
For instance, if all vector mesons are lighter than $2 \, m_\pi$, then two- and three-body visible decays are present in the theory.
The enhanced decay length from three-body processes favors high-current, long-baseline beam dump experiments for moderate values of $\epsilon$.\footnote{These three-body processes can also be used to test $SO(N_c)$ and $Sp(N_c)$ models, 
where the low-energy spectrum does not contain neutral pions, aside from the flavor singlet meson associated with the 
chiral anomaly~\cite{Hochberg:2015vrg}.
We leave a detailed study of these models to future work.}
A qualitatively different scenario arises if the $\omega$ and all charged vector mesons are heavier than $2 \, m_\pi$. Then, only two-body decays of 
the remaining neutral vector mesons can give rise to visible signatures, with a decay length better suited for short-
and medium-baseline experiments. We show in Sec.~\ref{sec:signals} and in Fig.~\ref{fig:EpsAndFpi} that both possibilities can be tested with existing data, future runs of HPS and SeaQuest, and the proposed LDMX experiment.
Alternatively, when $m_V > 2 \, m_\pi$, all vector mesons 
decay invisibly into pairs of DM pions. In this case, promising signals include DM-electron/nucleon scattering and missing energy signatures at accelerators~\cite{Hochberg:2017khi}.

\section{Cosmology\label{sec:cosmo}}
In this section, we discuss in more detail two aspects of 
SIMP cosmology mentioned in Secs.~\ref{sec:intro} and \ref{sec:model}. We previously noted 
that the singlet pions
are unstable even if $Q^2 \propto \mathbb{1}$ for general 
quark masses, $M_q$. If these are the lightest of the DM pions, then their 
decays can efficiently deplete the entire DM energy density.
Even if they are long-lived compared to the timescale of 
freeze-out, but with a lifetime greater than a second, late
decays may be in conflict with the successful predictions of Big Bang
nucleosynthesis or measurements of the CMB.

However, these pitfalls can be avoided. 
For example, if in addition to $Q^2 \propto \mathbbm{1}$, we enforce $\tr Q = 0$ and $M_q\propto \mathbb{1}$, then the singlet pions can be made absolutely stable through an enhanced symmetry. These requirements allow for the construction of a $G$-parity, 
\be
G \equiv C \times \mathbb{Z}_2^{\Ap} \times U_q
~,
\ee
where $C$ is the $U(1)_D$ charge conjugation operator, $\mathbb{Z}_2^{\Ap}$ sends $\Ap \to - \Ap$, and 
$U_q\in SU(N_f)_V$ is a unitary isospin transformation of the HS quarks such that $U_q^\dagger Q U_q = - Q^T$.
Note that such a $U_q$ exists only if the numbers of 
positively- and negatively-charged HS quarks are equal since a unitary transformation cannot alter the determinant or trace of $Q$. 
The singlet pion (defined in Eq.~(4.8) of Ref.~\cite{Hochberg:2015vrg})
is odd under $G$. Hence, the above choices of $Q$ and $M_q$ forbid the construction of an operator relevant for $\pi\rightarrow \Ap^{(*)}\Ap^{(*)} \to 4 \ell$ or $\pi \to 2 \ell$. This can easily be seen from the fact that any candidate amplitude must be built out of traces of at least two powers of $Q$, the generator 
corresponding to the pion, $T_\pi$, and possibly $M_q$. If $M_q\propto \mathbbm{1}$, these amplitudes can 
always be simplified to $\tr Q^2 \, T_\pi = 0$, $\tr Q ~ \tr Q \, T_\pi=0$, or $\tr Q^2 ~ \tr T_\pi = 0$.

\begin{figure}[t]
  \centering
\hspace{-0.5cm}
\includegraphics[width=5.5cm]{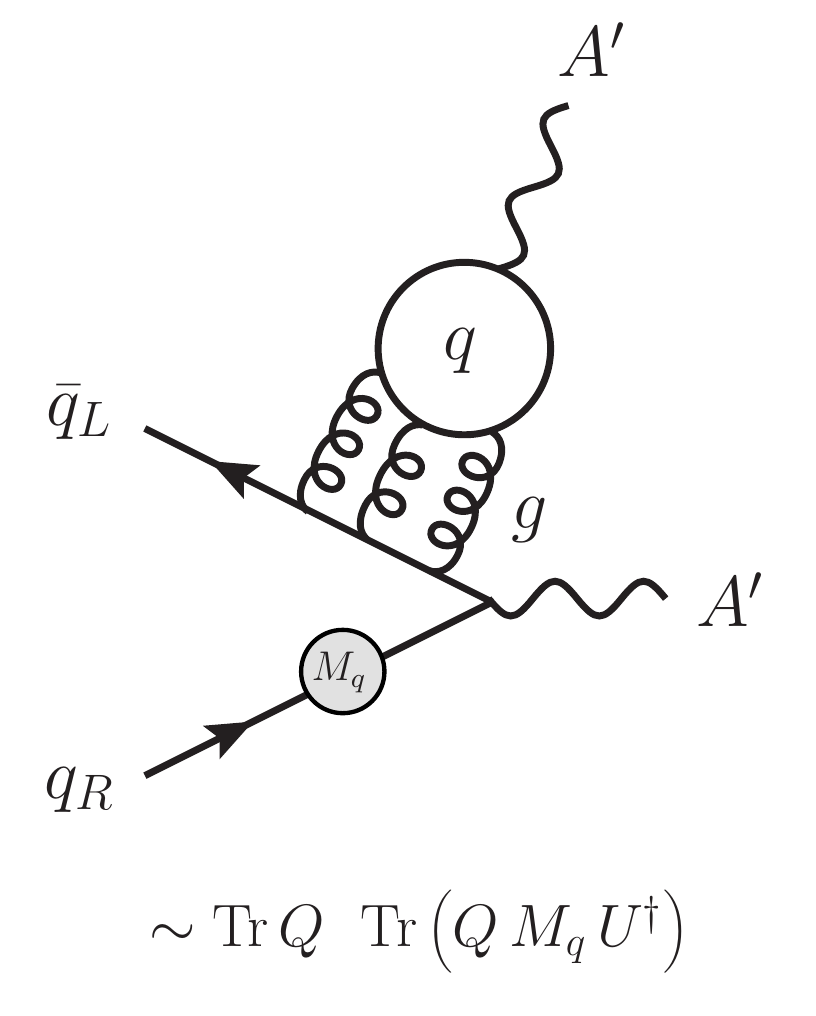} \hspace{-0.5cm}
\caption{An example of a hidden sector parton-level diagram and the corresponding operator in the chiral Lagrangian that are potentially responsible for the decay of singlet dark matter pions into Standard Model fermions. The gray circle represents an insertion of a hidden sector quark mass.} 
\label{fig:PionDecay}
\end{figure}

The situation is different if $N_f$ is odd or if the numbers of 
positively- and negatively-charged HS quarks are not equal. This is relevant for $N_f=3$ flavors as considered in this work.
In this case, even if $Q^2 \propto \mathbbm{1}$, there is no analogous 
$G$-parity, and there exist chirally-invariant operators that facilitate the decays of singlet 
pions, e.g.,
\be
\label{eq:piondecay1}
\frac{\alpha_D}{4 \pi  f_\pi} \, i \, \epsilon^{\mu\nu\alpha\beta} \, A^\prime_{\mu\nu} \, A^\prime_{\alpha\beta} \, \tr Q \, 
\tr \left( Q \, M_q \, U^\dagger \right) + \text{h.c.}
~,
\ee 
where $U = \exp(2 i \pi^a T^a/f_\pi)$, $T^a$ are the generators of $SU(N_f)$, and the overall coefficient is estimated using naive dimensional analysis~\cite{Georgi:1992dw}. At the parton level, traces of $Q^{2n-1}$ (with integer $n$) arise from quark loops with 
an odd number of $\Ap$ insertions. These diagrams have been claimed to vanish due to Furry's theorem~\cite{Hochberg:2015vrg}. 
However, this is not the case when an arbitrary number of HS gluons is attached to the loop. An example of such a parton-level diagram is shown in Fig.~\ref{fig:PionDecay}. 

\begin{figure*}[t]
  \centering
\includegraphics[width=12.0cm]{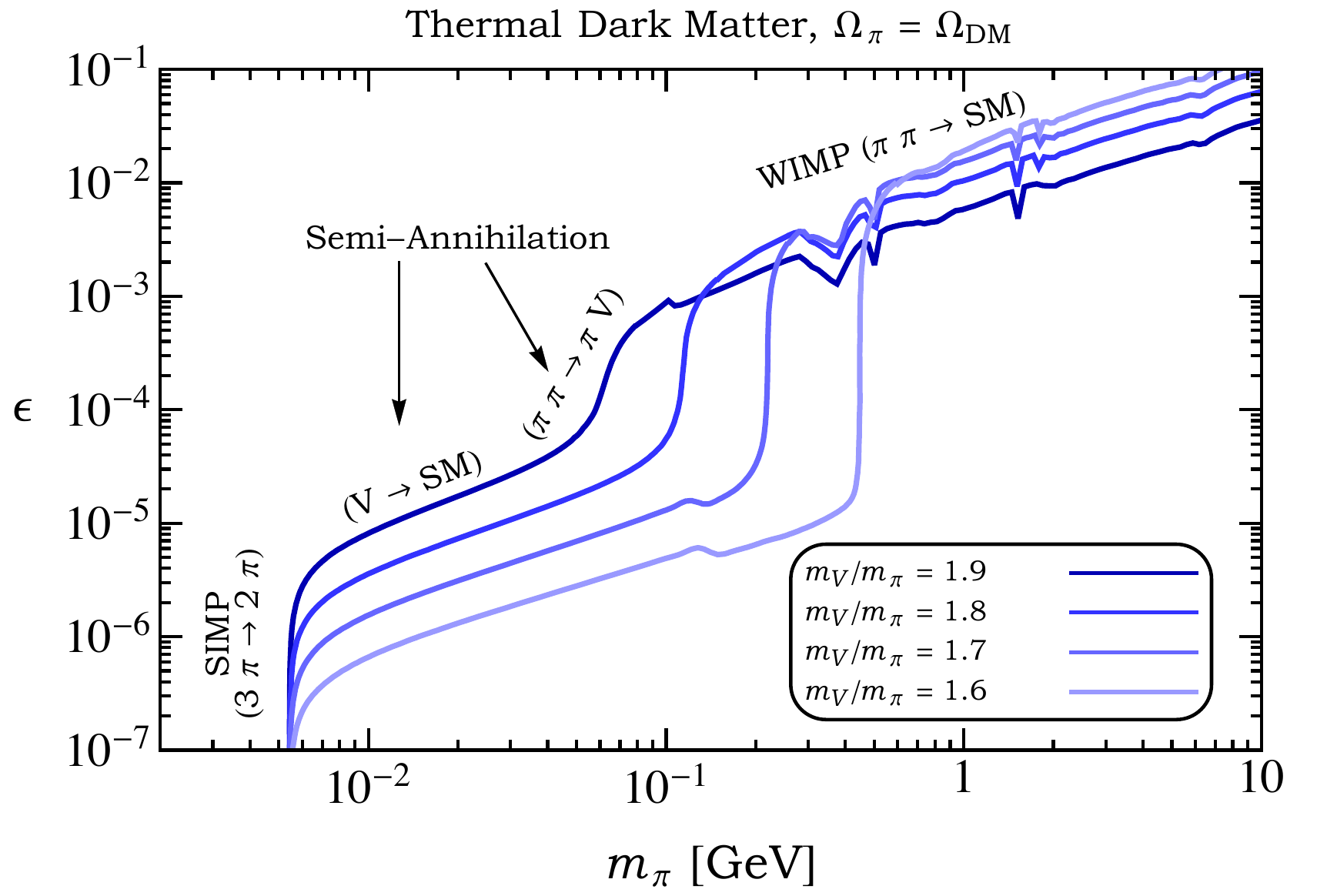} 
\caption{
  The hidden sector pion abundance in the $m_\pi - \epsilon$ plane. The contours correspond to regions where the $\pi$ abundance matches the observed dark matter energy density for $m_\pi/f_\pi = 3$, $\alpha_D = 10^{-2}$, $\mAp / m_\pi = 3$, and various values of the mass ratio, $m_V / m_\pi$. Multiple processes determine the dark matter 
  abundance. For $\epsilon \gtrsim 10^{-3}$, the abundance is governed by WIMP-like 
  annihilations into Standard Model particles, $\pi\pi \rightarrow \text{SM}$. For $\epsilon \sim 10^{-4} - 10^{-3}$, 
  efficient semi-annihilations, $\pi\pi\rightarrow \pi V$, dictate freeze-out. If $\epsilon \lesssim 10^{-5} - 10^{-4}$, 
  the ability of semi-annihilations to reduce the $\pi$ density becomes limited by the vector meson decay rate, $\Gamma (V\rightarrow \text{SM}) \sim \epsilon^2$. 
If $\epsilon \lesssim 10^{-6}$, then the relic abundance is determined by SIMP-like $3\rightarrow 2$ reactions. For simplicity, we have assumed that the hidden and visible sector baths remain in kinetic equilibrium throughout freeze-out. This assumption will be relaxed in Sec.~\ref{sec:signals}. Although this is valid for $\epsilon \gtrsim 10^{-6}$, we expect that kinetic decoupling will occur before dark matter freeze-out for $\epsilon \lesssim 10^{-6}$. This case is analogous to models in which the abundance is governed by kinetic decoupling~\cite{Kuflik:2017iqs,Kuflik:2015isi}. 
\label{fig:simp_abundance}}
\end{figure*}

The operator in Eq.~\eqref{eq:piondecay1} induces two- and four-body decays 
of the unstable singlet pions into SM fermions. Both partial widths are suppressed by 
$\alpha_D^2 \epsilon^4$ and by loop or phase space factors, leading to 
extremely long lifetimes. We estimate that the four-body channel dominates 
with a partial width of
\be
\Gamma (\pi \to 4 \ell) \sim \frac{\alpha_D^2 \, \alpha_\text{em}^2 \, \epsilon^4}{2048 \, \pi^5} ~ \frac{m_\pi^{11}}{f_\pi^2 \, \mAp^8}
~,
\ee
while the two-body channel is additionally suppressed by $m_\ell/m_\pi$.
For $\alpha_D = 10^{-2}$, $\epsilon\lesssim 10^{-4}$, and GeV-scale masses, the lifetime of the singlet pions is comparable to the timescale of recombination, $\sim 10^{13}$ seconds.
Measurements of the CMB anisotropy spectrum are a powerful probe of such an energy injection and, for certain lifetimes, limit
the fraction of DM that can undergo visible decays to $\lesssim 10^{-11}$~\cite{Slatyer:2012yq,Slatyer:2016qyl}.
Decays before recombination may be constrained by CMB spectral distortions~\cite{Slatyer:2016qyl}.
Thus, unless the abundance of the unstable component is negligible, the CMB places an important 
constraint on a wide swath of otherwise viable parameter space. We emphasize 
that this is not a generic feature of strongly interacting dark sectors since the HS pions 
can be made absolutely stable, as discussed above. 

Models where some of the pions are unstable may be cosmologically viable, provided all unstable species are heavier than the lightest stable pion and decay at temperatures below the corresponding mass splitting. In this case, $2 \to 2$ scattering in the HS depletes the abundance of the unstable pions before they decay.
CMB constraints on this possibility are discussed in Sec.~\ref{sec:CMB}. The relevant mass splittings can arise from chiral symmetry breaking corrections from HS quark masses, through higher-order operators in the chiral Lagrangian such as~\cite{Donoghue:1992dd}
\be
\alpha_{6,7} ~ B_0^2 ~ \left( \tr M_q \, U^\dagger \pm \text{h.c.} \right)^2
~,
\ee
where $B_0$ is a dimensionful constant related to the HS quark condensate and defined in Eq.~\eqref{eq:chiralLag1}. These operators, with coefficients $\alpha_{6,7} \sim \order{10^{-3}}$, can lift the unstable pions to the degree needed for cosmology if the values of $\alpha_{6,7}$ are chosen judiciously~\cite{Ecker:1988te,Donoghue:1992dd}. However, obtaining the cosmologically viable spectrum requires $\alpha_7 >0$ (of opposite sign from the expected $\eta'$-mixing contribution) and is about $2\sigma$ away from SM measurements~\cite{Donoghue:1992dd}. It is therefore unlikely (but not impossible) that this mechanism operates in SM-like theories. It is nonetheless plausible that this spectrum may be realized in less SM-like hidden sectors without $G$-parity.  

We now turn to the problem of DM freeze-out in the early universe. 
The first examples of SIMPs were designed to
realize the $3 \to 2$ annihilation mechanism~\cite{Hochberg:2014dra}. 
This process arises from a dimension-9 operator in the Wess-Zumino-Witten term. 
The corresponding rate in the early universe is given by
\be
\Gamma_{3\rightarrow 2} = \left(n_\pi^{\text{eq}}\right)^2 ~ \langle \sigma v^2 \rangle \propto 
\frac{ e^{-2x}}{x^5} ~ (m_\pi/f_\pi)^{10} ~ m_\pi 
~,
\label{eq:32_rate}
\ee
where $x \equiv m_\pi/T_{\mathrm{SM}}$, $T_{\mathrm{SM}}$ is the temperature of the SM bath, and $n_\pi^\text{eq}$ is the equilibrium number density of $\pi$.
Due to the strong exponential suppression in Eq.~\eqref{eq:32_rate},
the correct relic abundance is obtained for large values of $m_\pi/f_\pi$ close to the 
perturbativity bound, i.e., $m_\pi / f_\pi \sim 4\pi$~\cite{Hochberg:2014kqa}. This observation, 
combined with the effective field theory expectation for 
vector meson masses~\cite{Soldate:1989fh,Chivukula:1992gi,Georgi:1992dw},
\be
m_V \sim 4 \pi \, f_\pi / \sqrt{N_c}
~,
\label{eq:vector_meson_mass}
\ee
suggests that the 
vector mesons play an important role in the DM cosmology. Indeed, 
the Wess-Zumino-Witten term gives rise to the pion semi-annihilation ($2 \rightarrow 1$) process, $\pi\pi \rightarrow \pi V$, 
with $V$ decaying to the SM. This offers an alternative mechanism to deplete the pion number density.\footnote{In the SM, the same interaction 
is responsible for $\omega \rightarrow 3\pi$~\cite{Kaymakcalan:1983qq}.} 

We calculate the relic abundance of DM pions by numerically solving the full system of Boltzmann equations, incorporating  annihilations, $3 \to 2$ and $2 \to 1$ processes, and vector meson decays. A more detailed discussion is presented in Appendix~\ref{app:cosmo} and will also appear in forthcoming work. In Fig.~\ref{fig:simp_abundance}, we show regions in the $m_\pi-\epsilon$ plane where the $\pi$ abundance matches the observed DM energy density for $\alpha_D = 10^{-2}$, $m_\pi / f_\pi = 3$, $\mAp / m_\pi = 3$, and various values of $m_V / m_\pi$, assuming throughout that the HS and SM are in kinetic equilibrium. These parameter choices are motivated below.

The behavior of the relic abundance contours in Fig.~\ref{fig:simp_abundance} has a simple interpretation in terms of the dominant processes controlling freeze-out. The favored DM mass range is strongly dependent on $\epsilon$. At large $\epsilon$, direct annihilations into SM particles, $\pi \pi \to A^{\prime *} \to \text{SM}$, determine the $\pi$ relic abundance, favoring WIMP-like regions of parameter space. For smaller values of $\epsilon$, semi-annihilation ($\pi \pi \to \pi V$, $V \to \text{SM}$) dictates the cosmological history. If the vector mesons decay to SM particles more rapidly than they downscatter into HS pions, i.e., $\Gamma (V \to \text{SM}) \gtrsim n_\pi^\text{eq} \, \sigma v (\pi V \to \pi \pi)$, then semi-annihilation is independent of $\epsilon$ and is largely driven by the strength of $\pi \pi \to \pi V$ alone. DM freezes out in this case when
\be
\label{eq:fo1}
n_V^\text{eq} ~  \langle \sigma v (\pi V \to \pi \pi) \rangle \sim H
~,
\ee
where $H$ is the Hubble parameter. Alternatively, when $\Gamma (V \to \text{SM}) \lesssim n_\pi^\text{eq} \, \sigma v (\pi V \to \pi \pi)$, semi-annihilation is limited by the rate of vector meson decays, which depends on the strength of kinetic mixing, as in Eq.~\eqref{eq:Vdecay2body}. Hence, freeze-out occurs when 
\be
\label{eq:fo2}
\left(n_V^\text{eq} / n_\pi^\text{eq}\right) ~ \langle \Gamma (V \to \text{SM}) \rangle \sim H
~,
\ee
and the ability of semi-annihilation to deplete the pion number density is quenched by $\epsilon \ll 1$. Note that Eqs.~\eqref{eq:fo1} and~\eqref{eq:fo2} imply that semi-annihilation is dependent on $n_V^\text{eq} / n_\pi^\text{eq}$ and is exponentially sensitive to the mass ratio, $m_V / m_\pi$. For even smaller values of $\epsilon$, vector meson decay rates are extremely suppressed and semi-annihilation cannot efficiently deplete the pion abundance. At this point, the only remaining viable mechanism is the $3 \to 2$ process, $\pi \pi \pi \to \pi \pi$. This region of parameter space 
corresponds to standard SIMP models~\cite{Hochberg:2014dra}. 

In generating the contours of Fig.~\ref{fig:simp_abundance}, we have assumed that DM-SM elastic scattering and vector meson decays maintain kinetic equilibrium during freeze-out. This assumption breaks down in Fig.~\ref{fig:simp_abundance}   for $\epsilon \lesssim 10^{-6}$, in which case the HS and SM kinetically decouple \textit{before} $3\rightarrow 2$ reactions freeze out. This scenario is analogous to elastically decoupling relics (ELDERs) of Refs.~\cite{Kuflik:2015isi,Kuflik:2017iqs} but with a different process (vector meson decays to the SM) driving the kinetic equilibration of the two sectors.
We emphasize that for most of the viable parameter space, semi-annihilations govern the cosmology of DM freeze-out, while regions in which $3 \to 2$ processes or direct annihilations are important are mostly ruled out by existing constraints. This will be explored in more detail in Sec.~\ref{sec:signals}.
 
In the semi-annihilation region of Fig.~\ref{fig:simp_abundance}, the correct relic abundance 
is obtained for much larger values of $m_\pi$ compared to the pure $3 \to 2$ regime. 
Equivalently, smaller values of $m_\pi/f_\pi$ are feasible for fixed DM mass. This significantly alleviates tensions 
with perturbativity and bounds on DM self-interactions.\footnote{This was previously noted in Ref.~\cite{Kamada:2017tsq} 
in the context of semi-annihilation with an axion. 
While there is no compelling reason to have $m_{\mathrm{axion}} \simeq m_\pi$, 
HS vector mesons are naturally expected near this scale, as discussed at the beginning of Sec. \ref{sec:model}.}
We emphasize that the cosmological importance of processes involving vector mesons is a natural 
consequence of large values of $m_\pi/f_\pi \sim 4 \pi / (m_V / m_\pi)$ since $m_\pi / f_\pi \gg 1$ implies that $m_V \sim m_\pi$. If $m_V$ is fixed
according to Eq.~\eqref{eq:vector_meson_mass}, then the correct relic abundance can 
be obtained for a wide range of DM masses without conflicting with perturbativity or limits on DM self-interactions. For instance, if $\epsilon$ is 
large enough for semi-annihilations to be efficient (vector mesons rapidly decay to the SM) then Eq.~\eqref{eq:vector_meson_mass} implies that GeV-scale DM pions acquire an adequate relic density for\footnote{This form is motivated by a semi-analytic solution to the Boltzmann equations in Appendix~\ref{app:cosmo}, which will be presented in forthcoming work.}
\be
\frac{m_\pi}{f_\pi} \sim 3 \left( 1  + 0.1 \, \log{\frac{m_\pi}{10 \text{ MeV}}} \right)
~.
\ee
The mild dependence of $m_\pi/f_\pi$ on the DM mass motivates $m_\pi/f_\pi \simeq 3$ as a good representative choice for models 
of thermal DM.

Previous literature has focused on maintaining kinetic equilibrium through DM-SM elastic scattering via $\Ap$ exchange~\cite{Hochberg:2015vrg}. Since such processes are controlled by the strength of kinetic mixing, demanding that the DM and SM sectors remain in kinetic equilibrium during freeze-out places a lower limit on the size of $\epsilon$. The presence of neutral vector mesons provides another means of maintaining kinetic equilibrium between the two sectors. The decays and inverse-decays, $V \leftrightarrow \text{SM}$, efficiently equilibrate the SM and vector mesons, 
while $\pi\pi\leftrightarrow \pi V$ and $\pi V \leftrightarrow \pi V$ enforce equilibrium within the dark sector. 
For $m_V \lesssim 2 \, m_\pi$ and $m_\pi / f_\pi \lesssim 4 \pi$, equilibration between the HS and SM bath is dominantly governed by vector meson decays, which allows for significantly lower values of $\epsilon$. As mentioned above, for values of $\epsilon$ near this lower bound, kinetic decoupling of 
the HS from the SM occurs before $3 \to 2$ freeze-out, leading to ELDER-like cosmology, albeit at a much 
smaller $\epsilon$~\cite{Kuflik:2015isi,Kuflik:2017iqs}.

\section{Experimental Signals\label{sec:signals}}

\begin{figure*}[t]
  \centering
\hspace{-0.5cm}
\includegraphics[width=8.91cm]{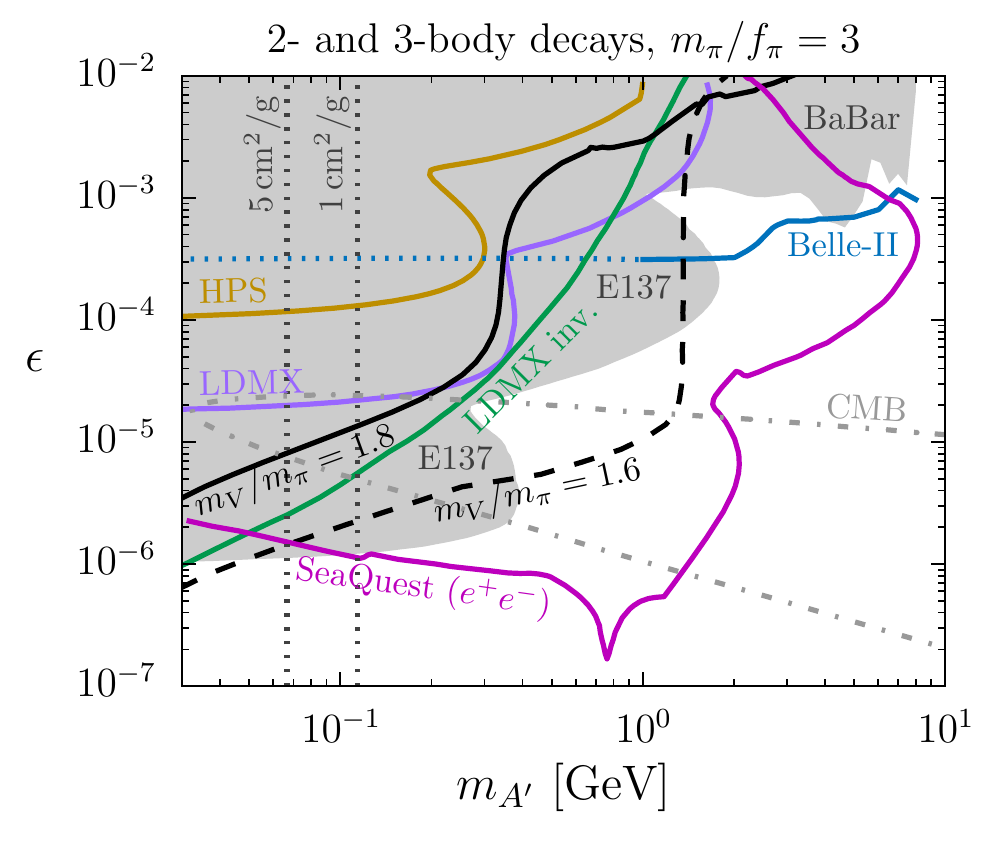} 
\includegraphics[width=8.91cm]{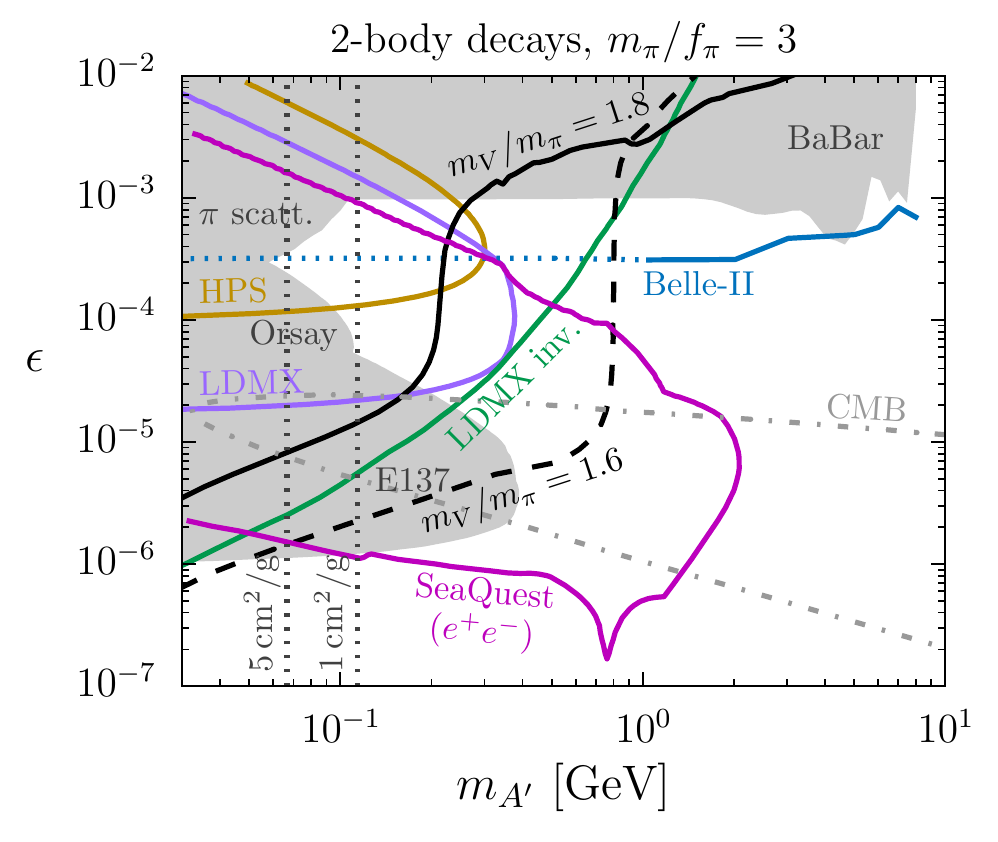} \hspace{0.5cm}
\includegraphics[width=8.91cm]{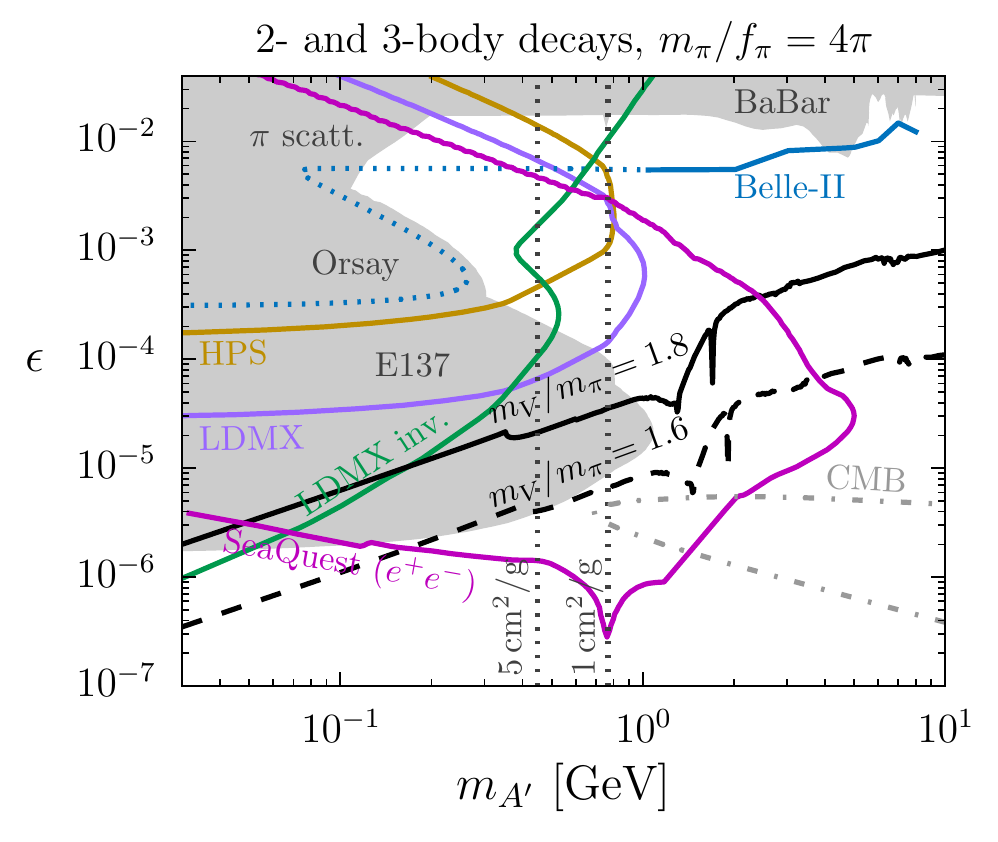} 
\includegraphics[width=8.91cm]{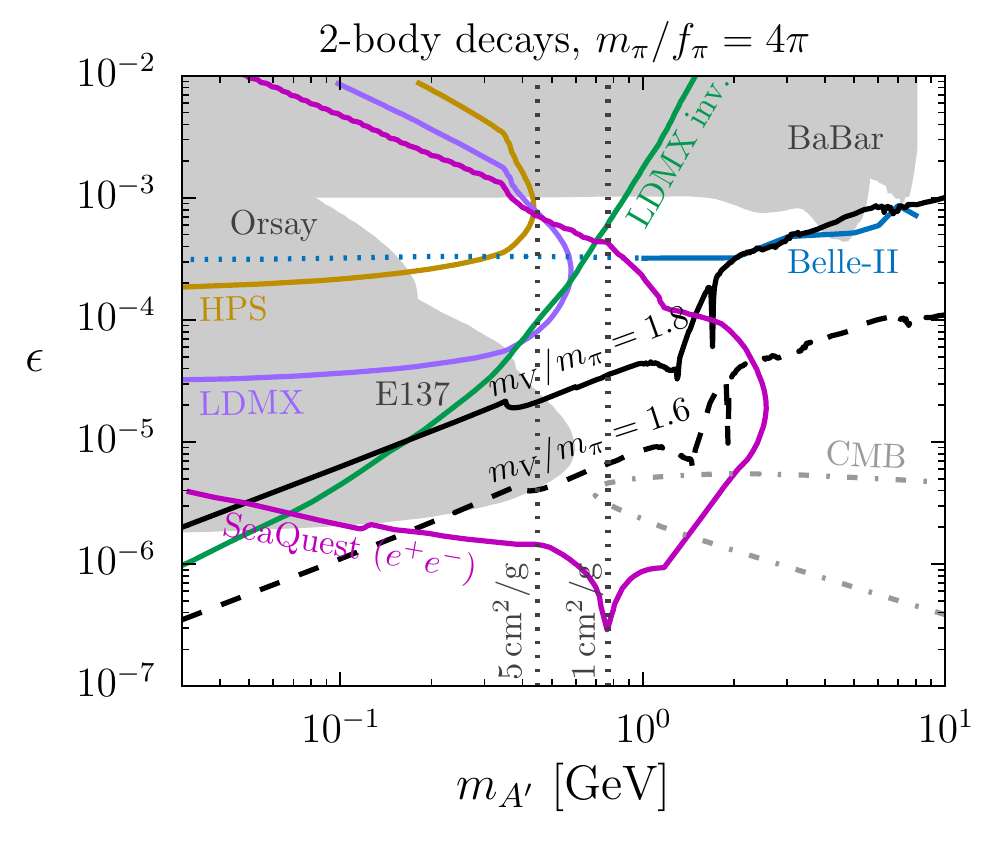} \hspace{0.5cm}
\caption{Existing constraints (gray regions) and sensitivity of future searches
  (colored lines) to signals of strongly interacting hidden sectors.  In the
  left column, we assume that all hidden sector vector mesons 
  decay to Standard Model particles via two-body ($V \rightarrow \ell^+ \ell^- $) or
  three-body ($V \rightarrow \pi \ell^+ \ell^-$) processes. In the right column, we
  assume that the vector mesons that do not mix with the dark photon are
  heavier than $2 \, m_\pi$. In this case, only the hidden sector $\rho$ and $\phi$ decay
  into Standard Model particles via two-body processes, while the remaining
  vector mesons decay invisibly into dark matter pions.   
  The shaded regions are
  excluded by BaBar~\cite{Lees:2017lec}, E137~\cite{Bjorken:1988as},
  Orsay~\cite{Davier:1989wz}, and searches for dark matter scattering at
  LSND~\cite{deNiverville:2011it}, E137~\cite{Batell:2014mga}, and
MiniBooNE~\cite{Aguilar-Arevalo:2017mqx}, as described
  in the text. The colored contours correspond to the projected reach of HPS~\cite{Celentano:2014wya}
  (orange), an upgraded version of SeaQuest~\cite{Aidala:2017ofy}
  (magenta), and the proposed LDMX experiment~\cite{Izaguirre:2014bca} (purple and
  green).  In evaluating experimental exclusions and
  projected sensitivity, we have fixed $\alpha_D = 10^{-2}$, $m_\Ap / m_\pi =
  3$, $m_V/m_\pi = 1.8$, and $m_\pi/f_\pi = 3$ ($4\pi$) in the top (bottom)
  row.   The experimental results are insensitive to small variations in
  $m_V/m_\pi$ (except for values near thresholds and resonances).
  In contrast, the dark matter abundance strongly depends on the $V$-$\pi$
  mass splitting (see Fig.~\ref{fig:simp_abundance}). In each panel, hidden
  sector pions make up all of the dark matter along the solid (dashed) black
  contours for $m_V/m_\pi = 1.8$ ($1.6$), while dark matter is overabundant
  below these lines. Even in the limiting case where $m_\pi/f_\pi = 4\pi$
  (which allows for the smallest coupling between the hidden sector and
  Standard Model), cosmologically favored regions of parameter space can be
  probed with existing and future experiments.  Contours of the dark matter 
  self-interaction cross-section per mass, $\sigma_\text{scatter} / m_\pi$, are shown as vertical gray dotted lines. For $N_f = \text{odd}$ flavors of light HS quarks, light dot-dashed gray contours denote regions excluded by measurements of the CMB. 
  } 
\label{fig:EpsAndFpi}
\end{figure*}

A rich experimental program is currently underway to search for light mediator
particles, such as the minimal dark photon, with beam dump, fixed-target, and
collider experiments~\cite{Alexander:2016aln,Battaglieri:2017aum}.  These
searches can also explore the strongly interacting models discussed in this
paper, which give rise to new signals from HS vector meson production and
decay, as shown in Fig.~\ref{fig:feyndiag}. In fact, existing data significantly constrains SIMP-like models.
However, much of the cosmologically motivated parameter space remains out of
reach. In this section, we show that searches for displaced visible decays of
HS vector mesons can be used to explore the best motivated regions
of masses and couplings. In Secs.~\ref{sec:E137}-\ref{sec:seaquest}, we focus
on existing data from past beam dump experiments and the
currently running HPS and SeaQuest experiments. In Sec.~\ref{sec:ldmx}, we
discuss signatures at the proposed LDMX detector. We also briefly comment on
possible searches at colliders such as BaBar, Belle-II, and the LHC in
Sec.~\ref{sec:collider}, the prospects for electron recoil signals at upcoming
low-threshold detectors in Sec.~\ref{sec:dd}, and bounds from galaxy clusters, supernova cooling, and the CMB in Secs.~\ref{sec:bullet}-\ref{sec:CMB}. Finally, in
Sec.~\ref{sec:expdisc}, we give a detailed discussion of our main results. These are summarized in Fig.~\ref{fig:EpsAndFpi}, where we show existing constraints and future projections as gray shaded regions
and colored contours, respectively.

\subsection{Past beam dumps}
\label{sec:E137}

The SLAC E137 beam dump experiment searched for long-lived particles 
produced in $e^-$-Al collisions. It utilized a $20 \GeV$ electron beam and $\sim 30 \text{ C}$ of current, equivalent to $\sim 10^{20}$ electrons on target (EOT) and $\sim 100 \text{ ab}^{-1}$ of integrated luminosity~\cite{Bjorken:1988as}. 
An electromagnetic calorimeter (ECAL) was placed $\sim 400$ meters downstream of the target 
after $\sim 200$ meters of natural shielding provided by a dirt hill and a $\sim 200$ meter open air decay region.
No signal events were observed.  At this experiment, HS vector mesons may have been produced as shown in Fig.~\ref{fig:feyndiag}. Following Refs.~\cite{Bjorken:2009mm,Andreas:2012mt}, we use \texttt{MadGraph5}~\cite{Alwall:2014hca} to calculate the detector acceptance and event yield. 

If a HS vector meson is sufficiently long-lived, 
it can traverse the hill before decaying into electrons in the open region. As shown in the lower panel of Fig.~\ref{fig:ApBF} for $m_\pi/f_\pi=3$, the lifetime of HS vector mesons that decay to the SM through three-body (two-body) processes is optimal for E137 for $\epsilon \sim 10^{-4} - 10^{-3}$ $(10^{-5} - 10^{-6})$ and GeV-scale vectors. Data from E137 is sensitive to such events as long as these electrons are energetic and directly pass through the ECAL. We define the signal region by requiring that an electron from such decays points back to the target within a resolution of $\sim \order{10} \text{ mrad}$ and deposits more than $1-3$ GeV of energy in the calorimeter. This range of energies corresponds to the different analysis thresholds utilized in Ref.~\cite{Bjorken:1988as}. We also include the effect of energy loss as the electron traverses the open air, incorporating a radiation length of $\sim 304 \text{ meters}$~\cite{Patrignani:2016xqp,Tsai:1966js}. This effect is typically negligible in searches for minimal dark photons or axions. However, in the models studied here, final state leptons are produced from a cascade of decays and inherit a smaller fraction of the initial beam energy. For comparable $\pi$ and $V$ masses, 
$e^\pm$ from the decay $V \to e^+ e^-$ each carry only about $\Ebeam/4 \simeq 5 \GeV$ of energy. These electrons then lose $\sim 50\%$ of their energy on average 
while traversing the air, making the location of the decay and 
the precise value of the energy threshold important. These effects are more significant for three-body decays of HS vector mesons, as the final state electrons are comparatively softer. For $\mAp \lesssim \text{GeV}$, roughly $\gtrsim 90 \%$ and $\gtrsim 5 \% - 90\%$ of events pass the pointing and energy threshold cuts, respectively, where the lower part of the range corresponds to demanding a minimum energy deposition of $3 \GeV$ from $V^\pm \to \pi^\pm e^+ e^-$. For concreteness, we choose a 1 GeV threshold.\footnote{For the models studied here, the E137 reach does not appreciably change for a threshold of 3 GeV.}  The E137 regions in Fig.~\ref{fig:EpsAndFpi} correspond to $95\%$ C.L. exclusions where at least $3$ signal events are predicted.

Long-lived particles can also be produced at proton beam dumps~\cite{Batell:2009di}. Experiments like CHARM~\cite{Gninenko:2012eq} and 
$\nu$-Cal~\cite{Blumlein:2011mv} are sensitive to a similar region of parameter space as E137~\cite{Bjorken:2009mm}.
Shorter baseline beam dumps, such as the Orsay experiment~\cite{Davier:1989wz}, offer 
complementary sensitivity to smaller decay lengths. 
For $\mAp \lesssim \order{100} \, \MeV$, decays of the $\Ap$
into DM pions are constrained by searches for $\Ap$ production and decay followed by
DM scattering in a downstream detector such as at
LSND~\cite{deNiverville:2011it,Auerbach:2001wg},
E137~\cite{Batell:2014mga,Bjorken:1988as}, and
MiniBooNE~\cite{Aguilar-Arevalo:2017mqx}. These regions are labeled as ``$\pi$-scatt.'' in Fig. \ref{fig:EpsAndFpi}.

\subsection{HPS}
\label{sec:HPS}

\begin{figure*}[t]
  \centering
\hspace{-0.5cm}
\includegraphics[width=8.91cm]{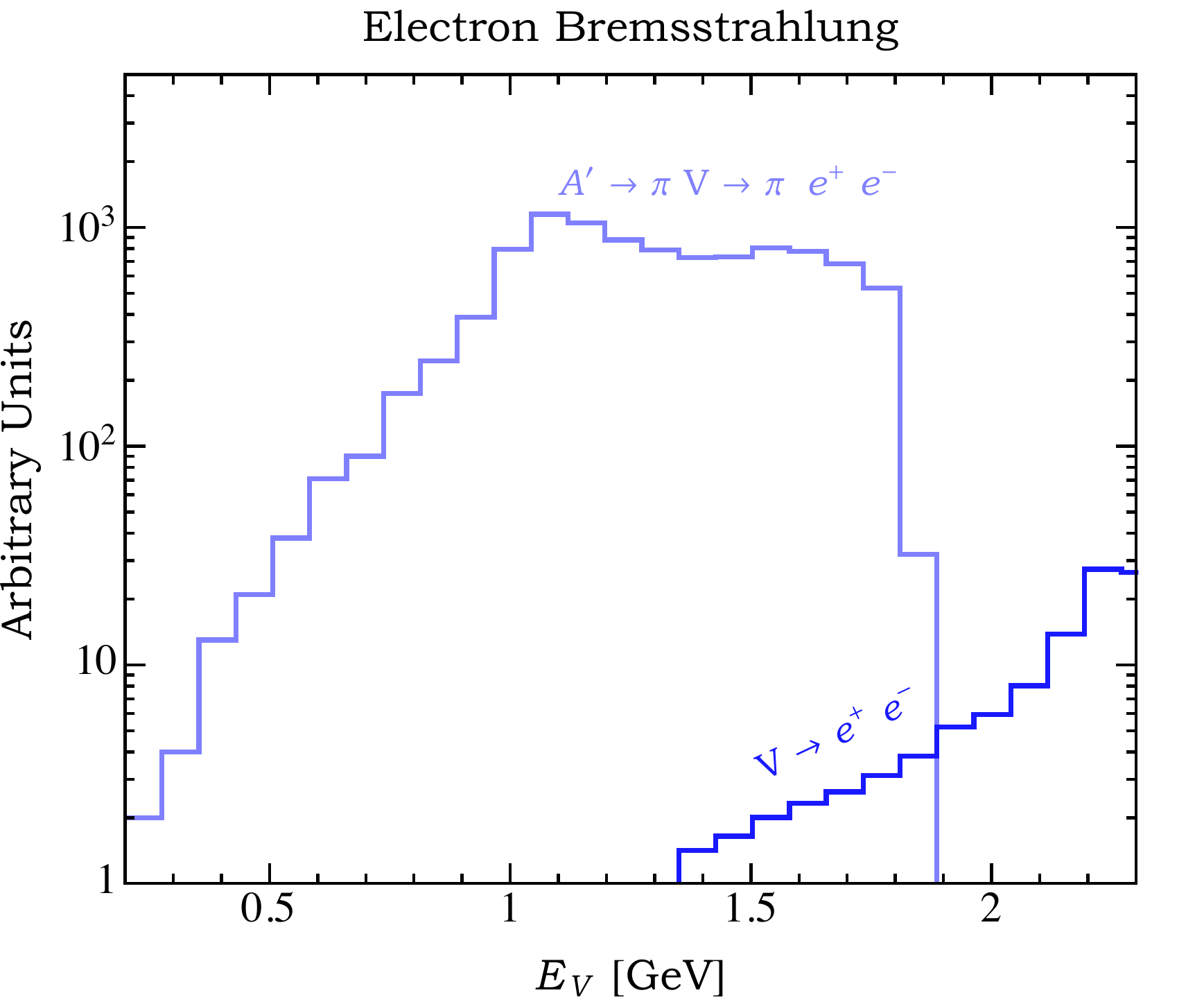} 
\includegraphics[width=8.91cm]{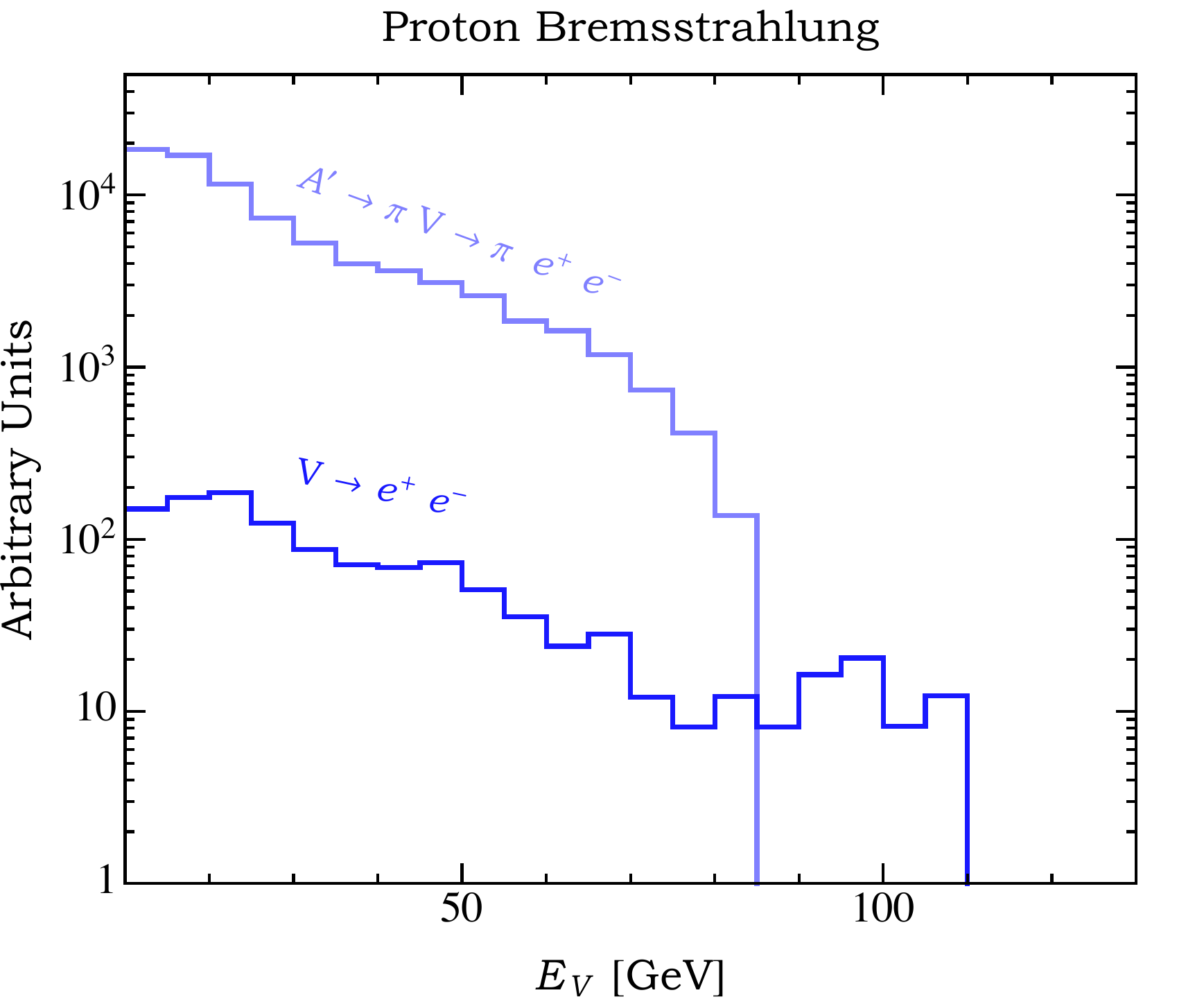} \hspace{0.5cm}
\caption{Signal kinematics of electron (left) and proton (right) Bremsstrahlung for $m_\pi = 100\text{ MeV}$, $m_V = 180 \text{ MeV}$, $\mAp = 300 \text{ MeV}$, $\alpha_D = 10^{-2}$, and $m_\pi / f_\pi = 3$. We show the distribution in the energy of the visibly decaying hidden sector vector meson, $E_V = E_{e^+ e^-}$, for both direct ($V \to e^+ e^-$) and indirect production through the decay of an on-shell dark photon ($\Ap \to \pi V \to \pi e^+ e^-$), the latter of which is the focus of this work. In the left (right) panel, the beam energy and decay length are fixed to those of the HPS (SeaQuest) experiment. In the right panel, we have additionally selected events that pass the geometric criteria as detailed in Sec.~\ref{sec:seaquest}. 
Direct $V$-Bremsstrahlung is suppressed by $V-\Ap$ mixing relative to production through $\Ap$ decay. As a result, 
for a fixed $V$ lifetime, the indirect $V$-production mechanism dominates, leading to enhanced displaced vertex rates relative to 
the minimal dark photon-like scenario.
}
\label{fig:kin}
\end{figure*}

The Heavy
Photon Search (HPS) experiment is well-suited to detect visible decays of the HS vector mesons~\cite{Celentano:2014wya}.  The first
science run was completed in 2016 with a
$2.3 \GeV$ electron beam, a $4\;\mu\mathrm{m}$ tungsten target, and $5.6\times 10^{17}$ EOT. 
A future run is anticipated with a more energetic beam ($6.6 \GeV$), a thicker target ($8.8 ~ \mu \text{m}$), and a larger luminosity ($
6.8\times 10^{18}$ EOT)~\cite{JJ:2017ab}. We simulate vector meson production (shown in Fig.~\ref{fig:feyndiag}) using \texttt{MadGraph5}~\cite{Alwall:2014hca}. The HPS spectrometer is a small-acceptance detector. We demand that the vector mesons decay to $e^+e^-$ pairs within $1-8$ cm of the target and assume that the geometric and kinematic efficiency is $5 -10 \%$.\footnote{We thank Takashi Maruyama
  and Bradley Yale for evaluating this acceptance.} 
The displaced vertex, along with the $e^+e^-$ invariant mass peak (from $V \to e^+ e^-$) and missing energy (due to the HS pions) can be used to suppress backgrounds to negligible levels. The 
future sensitivity of HPS, corresponding to a yield of $\sim 5 -10$ events, is shown as the orange contours in Fig.~\ref{fig:EpsAndFpi}. The kinematic distribution of this signal is shown in the left panel of Fig.~\ref{fig:kin}. 

\subsection{SeaQuest}
\label{sec:seaquest}

The SeaQuest experiment is currently running at the Fermi National Accelerator Laboratory (FNAL)~\cite{Aidala:2017ofy}. It is a nuclear physics spectrometer primarily designed to measure SM Drell-Yan production of muons. The detector trigger has recently been upgraded to retain both low-mass and long-lived dimuon pairs~\cite{doi:10.1142/S0217732317300087,mingmaryland,Sanghoon}. Future plans also include the installation of a recycled ECAL, which would allow for additional sensitivity to displaced electrons. SeaQuest utilizes the FNAL 120 GeV main injector proton beam and consists of a 5 meter magnetized iron target/shield (FMAG), a 3 meter focusing magnet (KMAG), and a series of triggering/tracking stations designed for accurate vertex and momentum reconstruction. A parasitic implementation of the displaced vertex trigger is expected to acquire $1.44 \times 10^{18}$ protons on target (POT) within the next two years, and a more dedicated run could obtain an integrated luminosity of up to $10^{20}$ POT~\cite{TalkStatusSeaQuest,Shiltsev:2017mle}.

We consider three mechanisms for dark photon production at SeaQuest:
Bremsstrahlung in the proton-nucleus collisions (see Fig.~\ref{fig:feyndiag}),
Drell-Yan, and SM meson decays, e.g., $\pi^0_\text{SM} \to \gamma
\Ap$~\cite{Gardner:2015wea}. 
The kinematic distribution of the signal from proton Bremsstrahlung is shown in the right panel of Fig.~\ref{fig:kin}. 
As in the previous sections, the dark photons decay promptly into HS vector mesons ($\Ap \to \pi V$).
For sufficiently small $\epsilon$, $V$ can easily traverse the iron shield before
decaying into a lepton pair in the region downstream of FMAG. If $V$ decays to
$e^+ e^-$, the level of SM backgrounds is expected to be small, as the magnetic
field of FMAG effectively sweeps away soft SM radiation. The optimal ECAL
location and form of the displaced electron trigger are uncertain. We will
therefore investigate the conservative scenario in which decays occur
immediately after FMAG and before the first existing tracking station,
corresponding to a decay length of $\sim 5 - 6$ meters (as measured from the front of FMAG). We simulate production
and decays at SeaQuest using \texttt{MadGraph5}~\cite{Alwall:2014hca} and {\tt PYTHIA 8.2}~\cite{Sjostrand:2014zea}. 
We estimate the reach of a future displaced electron search by demanding at least
10 events in which $V$ decays to $e^+ e^-$ in the fiducial region of $5-6$
meters. We further require that the electron pair remains within the geometric
acceptance of the spectrometer, incorporating the transverse momentum kick of
KMAG, $|\Delta p_T| \simeq 0.4 \GeV$. The magenta contours in
Fig.~\ref{fig:EpsAndFpi} illustrate the potential capability of a future
version of SeaQuest, with an upgraded sensitivity to electrons and $10^{20}$
POT. More details on the modeling and capability of this instrument
will be presented in the forthcoming work, Ref.~\cite{Berlin:2018pwi}. We also
note that similar sensitivity and an enhanced mass reach could be achieved with
the proposed SHiP experiment (which utilizes the 400 GeV proton beam at CERN SPS)
for dark photons heavier than 10 GeV~\cite{Alekhin:2015byh}.

\subsection{LDMX}
\label{sec:ldmx}
 
The proposed LDMX experiment will be able to test SIMP-like models in two distinct
channels: visible displaced $V$ decays and invisible dark photon decays. 
LDMX is designed to look
for missing momentum by accurately measuring a beam electron's 
momentum before and after it scatters in a $\sim 10\%$ radiation-length tungsten target~\cite{Izaguirre:2014bca,Alexander:2016aln,Battaglieri:2017aum}. 
Two experimental phases are planned. 
Phase 1 (2) will acquire $4\times 10^{14}$ ($10^{16}$) EOT with a beam energy of $\Ebeam = 4$ (8) GeV.

The missing momentum program at LDMX requires a large hermetic calorimeter in
order to reliably measure the scattered electron energy. This opens the
possibility of performing a search for the same visible signals discussed in the previous sections. 
If a HS vector meson produced in the collision decays to $e^+ e^-$
sufficiently far inside the calorimeter, it can give rise to a displaced electromagnetic
shower. The size of the detector can then be used to range out normal electromagnetic
backgrounds, such as the shower from the scattered electron. 
The majority of signal events originate from $\Ap$ production in the first radiation length of the ECAL and not from the thin target.
We model this signal as described in Sec.~\ref{sec:HPS} and estimate 
the sensitivity of such a search using the Phase 2 luminosity
by requiring that the visible decays occur 
between $7$ cm (20 radiation lengths in tungsten) and 300 cm (approximate length of the detector). 
We select candidate events such that the recoil electron energy is less than $0.3 \, \Ebeam$ 
and apply an overall 50\% ``quality'' cut to the yield to crudely model detector efficiencies. 
The estimated sensitivity, corresponding to a yield of 3 events, is shown as the purple contours in Fig.~\ref{fig:EpsAndFpi}.

The other class of signals arises from invisible $\Ap$ decays. For the SIMP models discussed here, this signal occurs when the $\Ap$ decays to 
$\pi\pi$ or to vector mesons that decay outside of the detector.
As shown in Fig.~\ref{fig:ApBF}, the branching fraction for $\Ap\rightarrow \pi\pi$
is $\mathcal{O}(10) \%$ for cosmologically motivated parameters, $m_\pi/f_\pi \lesssim 5$, and 
falls rapidly for larger $m_\pi/f_\pi$.
We estimate the signal yield for the missing momentum search by rescaling the results 
in Refs.~\cite{Alexander:2016aln,Battaglieri:2017aum}, including decays to HS pions and vector mesons. 
The resulting reach is shown as the green contours in Fig.~\ref{fig:EpsAndFpi}.
The currently running NA64 experiment is sensitive to similar missing mass phenomenology~\cite{Banerjee:2016tad,Banerjee:2017hhz}. 
However, the present dataset does not probe parameter space beyond existing beam dump constraints.

\subsection{Colliders}
\label{sec:collider}
Processes analogous to those presented in Fig.~\ref{fig:feyndiag} can be looked for at low-energy 
$e^+e^-$ colliders and at the LHC.
Searches at BaBar for invisible decays of dark photons in association with a SM photon
constitute the strongest probe of strongly interacting hidden sectors 
at present for $1 \GeV \lesssim \mAp \lesssim 10 \GeV$~\cite{Lees:2017lec,Aubert:2008as}.
In recasting this search, we have included decays of the $\Ap$ into HS pions and long-lived
vector mesons, demanding that the latter decay outside of the detector (modeled as a cylinder with a 1.5 meter radius). A
similar search at Belle-II is expected to improve these limits in the near
future, probing values of $\epsilon$ up to a factor of $\sim 3$ smaller
compared to BaBar with only $20$ $\text{fb}^{-1}$ of
data~\cite{Battaglieri:2017aum}. 
In Fig.~\ref{fig:EpsAndFpi}, the Belle-II reach is shown as a solid (dotted) blue contour for $\mAp \gtrsim \text{GeV}$ ($\mAp \lesssim \text{GeV}$). We note that this sensitivity projection for $\mAp \lesssim \text{GeV}$ is 
relatively uncertain as it relies on the rejection and measurement
of a peaking background~\cite{Essig:2013vha}. 

Searches for new GeV-scale particles at the LHC are complicated by enormous SM backgrounds, which 
can be suppressed by looking for, e.g., resonances and displaced vertices.
Refs.~\cite{Ilten:2015hya,Ilten:2016tkc} used these techniques to demonstrate that
LHCb will have impressive sensitivity to the minimal visible dark photon decays ($\Ap \to \ell^+ \ell^-$) in Run 3 of the LHC. 
In fact, an analysis has recently been performed with limited data in Ref.~\cite{Aaij:2017rft}. 
These studies rely on kinematic features that are unique to $\Ap\rightarrow \ell^+ \ell^-$
 in key ways. For example, the $D^*\rightarrow D^0 \Ap$ search
requires the lepton pair and $D^0$ momenta to reconstruct the $D^*$; this is not the case for the
cascade decays considered in this work, $\Ap \rightarrow \pi V (\to \ell^+ \ell^-)$, 
due to the presence of the final state HS pions~\cite{Ilten:2015hya}. The inclusive analysis of
Ref.~\cite{Ilten:2016tkc} relies on data-driven estimates of signal and
therefore it is not simple to determine the signal and reach to the more complicated $\Ap$ decays
considered here (for recent work on recasting LHCb dilepton searches, see Ref.~\cite{Ilten:2018crw}). It would be interesting to investigate 
if the above search strategies can be generalized to signals present in SIMP-like theories.

Other, more inclusive, searches at $B$-factories and the LHC can also be used to study the 
hidden sectors discussed here.
For instance, BaBar has performed a search for pairs
of tracks that reconstruct a displaced vertex, which may arise from the
leptonic or hadronic decays of long-lived particles~\cite{Lees:2015rxq}. 
For momentum transfers below the HS confinement scale, these types of signals
arise from the same visible decays of HS vector mesons, e.g., $e^+ e^- \to \pi
V (\to \mu^+ \mu^-) + \cdots$. Processes involving larger momentum transfers
must be treated as production of HS quarks~\cite{Strassler:2006im} that
subsequently shower and hadronize, yielding a large multiplicity of pions and
vector mesons~\cite{Cohen:2015toa,Pierce:2017taw,Cohen:2017pzm,Park:2017rfb}. At the LHC, similar searches
at (or near) the ATLAS, CMS, and LHCb detectors could also offer greater
sensitivity to heavier HS states, leveraging the additional presence of
missing energy arising from the production of stable DM
pions or the long lifetimes of HS vector mesons~\cite{Izaguirre:2015zva,Chou:2016lxi,Feng:2017uoz,Gligorov:2017nwh}. 
We leave a detailed exploration of
these signals to future work.

Throughout this work, we have assumed that the SM-HS mediator can be produced on-shell.
If the dark photon is above the kinematic threshold of certain 
accelerators, HS states can be produced directly 
through an off-shell $\Ap$.
Standard constraints on visibly decaying dark photons can then be applied to the HS vector mesons, 
whose coupling to SM states is described by an effective kinetic mixing, $\epsilon_\text{eff} \sim \epsilon\, (e_D\,/\,g)\,(m_V^2 \, / \, \mAp^2)$,
generated by a heavy $\Ap$.
For instance, taking $\mAp / m_V \simeq 100$ and $\alpha_D \sim 10^{-1}$, constraints from 
long-baseline beam dumps, which are typically sensitive to $\epsilon \sim 10^{-8} - 10^{-6}$ for light visibly decaying dark photons~\cite{Alexander:2016aln,Battaglieri:2017aum}, instead probe $\epsilon \sim 10^{-4} - 10^{-2}$ when $V$'s are produced through an off-shell $\Ap$. 
However, these limits are always subdominant to the ones discussed throughout this work when on-shell dark photons can be produced and decay to long-lived vector mesons.

\subsection{Direct Detection}
\label{sec:dd}

Light DM can also be detected with electron or nuclear recoils in 
low-threshold detectors~\cite{Essig:2015cda,Battaglieri:2017aum}. If the global $U(1)_D$ symmetry is 
preserved by the HS quark masses, i.e., $[Q,M_q] = 0$, then $\Ap$ exchange 
can lead to elastic scattering in semi-conductor direct detection experiments 
such as SENSEI and SuperCDMS~\cite{Battaglieri:2017aum}. Using the optimistic thresholds and exposures from Ref.~\cite{Essig:2015cda}, we find that SENSEI will be able to test regions of parameter space in the upper-left corner of each panel in Fig. \ref{fig:EpsAndFpi} (see also Ref. \cite{Hochberg:2015vrg}).
Direct detection signals are absent if the $\Ap$ couples off-diagonally to 
mass eigenstates that are split by more than $ m_\pi v^2 \sim  0.05 \,\keV\, (m_\pi/50\,\MeV)$.
This can occur if there is a single Higgs field that is responsible for both the $\Ap$ and HS quark masses, resulting in $[Q,M_q]\neq 0$.
If such mass splittings are smaller than the temperature at DM freeze-out, $T \sim m_\pi/20$, the cosmological evolution, 
as well as signals at fixed-target and collider experiments, are not significantly affected.

\subsection{Self-Interactions}
\label{sec:bullet}

The self-scattering cross-section per DM mass is bounded from observations of galaxy clusters, $\sigma_\text{scatter} / m_\pi \lesssim \text{few} \times \text{cm}^2/\text{g}$~\cite{Clowe:2003tk,Markevitch:2003at,Randall:2007ph,Kaplinghat:2015aga}. As in Ref.~\cite{Hochberg:2014kqa}, we calculate this bound assuming that a single species of pion is the lightest state in the theory. As a representative benchmark, we show contours of $1$ and $5 \text{ cm}^2/\text{g}$ as vertical gray dotted lines in Fig.~\ref{fig:EpsAndFpi}. This bound does not apply if the HS pions make up a subdominant fraction ($\lesssim 10 \%$) of the total DM energy density.

\subsection{Supernovae}
\label{sec:SN}

Light particles that are feebly coupled to the SM can lead to anomalous cooling in stellar systems. For instance, DM or long-lived states that are produced in supernovae from an off-shell intermediate $\Ap$ can diffuse out of the core, leading to energy loss. For $m_{\pi, V} \lesssim 10 \text{ MeV}$, such processes could have lead to qualitative changes in the cooling rate and neutrino burst duration of supernova SN1987A. From the discussion in Ref.~\cite{Izaguirre:2014bca}, we estimate that these bounds may apply for couplings of size $\alpha_D \, \epsilon^2 \sim 10^{-15}$. Although not shown, this potential exclusion corresponds to the region of parameter space in the lower-left corner of each panel in Fig.~\ref{fig:EpsAndFpi}. 
A careful determination of the supernova cooling bound will appear in Refs.~\cite{Chang:2018rso,Graham:2018}.

\subsection{CMB}
\label{sec:CMB}

For $N_f = \text{odd}$ flavors of light HS quarks, the singlet DM pions are cosmologically unstable. As discussed in Sec.~\ref{sec:cosmo}, reasonable coefficients for higher-order operators in the chiral Lagrangian lift the masses of the unstable pions above those of the stable species. In this case, $2 \to 2$ processes (such as $\pi^0 \pi^0 \to \pi^\pm \pi^\pm$) significantly deplete the fractional abundance of the unstable pions before they decay to SM particles. 

Regardless, measurements of the CMB are extremely sensitive to energy injections at the time of recombination from a decaying DM subcomponent~\cite{Slatyer:2012yq,Slatyer:2016qyl}. In estimating this bound, we take the limits from Ref.~\cite{Slatyer:2016qyl} and assume that the unstable and stable pion species are split in mass at the level of $\Delta \sim 5\%$. This choice does not significantly affect the freeze-out calculations discussed throughout this work since DM chemically decouples at $T \sim m_\pi / 20$. The frozen-out fractional abundance of the unstable species is then proportional to $\exp(-\Delta \, x_{2 \to 2})$, where $x_{2 \to 2} \sim \order{100}$ is estimated by demanding that $n_\pi \, \sigma v (2 \to 2) \sim H$ when $T_\text{SM} \sim m_\pi / x_{2 \to 2}$. 

Regions of parameter space that are in conflict with measurements of the CMB anisotropies are shown as the light gray dot-dashed contours in Fig.~\ref{fig:EpsAndFpi}. This constraint is severely weakened for larger values of $m_\pi / f_\pi$ since $2 \to 2$ scattering more efficiently depopulates the unstable species. We emphasize that this exclusion is not intrinsic to models of strongly interacting DM. For instance, as discussed in Sec.~\ref{sec:cosmo}, the DM pions can be made absolutely stable for $N_f = 4$ light flavors, while considerations of DM freeze-out and the phenomenology in Fig.~\ref{fig:EpsAndFpi} are not qualitatively affected.

\subsection{Discussion of Results}
\label{sec:expdisc}

We show existing constraints and the projected reach of
experiments in Fig.~\ref{fig:EpsAndFpi} as gray regions and
colored contours, respectively, in the $\mAp - \epsilon$ plane. In each panel, we have fixed $\alpha_D =
10^{-2}$, $m_V / m_\pi = 1.8$, and $\mAp / m_\pi = 3$. In the top (bottom)
row, we take $m_\pi / f_\pi = 3$ ($4 \pi$). Since
we do not expect small variations in the vector meson-pion mass ratio, $m_V / m_\pi$, to
significantly impact the phenomenology, thermal freeze-out targets are shown for
both $m_V / m_\pi = 1.8$ and $m_V / m_\pi = 1.6$ as solid and dashed black
lines, respectively. The final abundance of the HS pions
matches the observed DM energy density along these contours.

In the left panel of each row of Fig.~\ref{fig:EpsAndFpi}, we assume that both the neutral and charged vector mesons are lighter than $2 \, m_\pi$. In this case, both two- and three-body decays to SM particles are present in the theory, corresponding to the processes shown in the bottom-left and bottom-right diagrams of Fig.~\ref{fig:feyndiag}. The large $\Ap$ production rate makes E137 a powerful test of this mass spectrum. As shown in Fig.~\ref{fig:EpsAndFpi}, E137 excludes a wide swath of models in which sub-GeV HS pions make up all of the DM. 
In this model, beam dump and fixed-target experiments are sensitive to two distinct ranges of $\epsilon$, due to the differing lifetimes of vector mesons that undergo two- or three-body decays (see Fig.~\ref{fig:ApBF}). For $m_\pi / f_\pi = 3$, this results in the two lobe-like features in the top-left panel of Fig.~\ref{fig:EpsAndFpi}, corresponding to two-body (lower lobe) and three-body (upper lobe) vector meson decays.

A qualitatively different picture emerges if only two-body decays of vector mesons into the SM are allowed, as in the right panel of each row of Fig.~\ref{fig:EpsAndFpi}. This scenario arises if only the vector mesons that directly mix with the $\Ap$ are lighter than $2 \, m_\pi$. The heavier vector mesons decay invisibly to pairs of HS pions, and the 
process shown in the bottom-right diagram of Fig.~\ref{fig:feyndiag} does not occur with an appreciable rate. For this mass spectrum and $\epsilon \gtrsim 10^{-4}$, the decay lengths of vector mesons are much shorter than the E137 baseline, leaving open a large set of viable models that can be explored in the near future. The discovery prospects of HPS, SeaQuest, and LDMX are shown as the colored contours in
Fig.~\ref{fig:EpsAndFpi}. HPS can probe cosmologically interesting models in a future run given the mass spectrum shown in the right panels. 
On longer timescales, LDMX and especially SeaQuest will explore new territory. 

For larger values of $m_\pi / f_\pi$, as shown in the bottom row of
Fig.~\ref{fig:EpsAndFpi}, the branching ratio for $\Ap \to V\pi$ is enhanced
(see Fig.~\ref{fig:ApBF}). More importantly, the decay lengths corresponding to
$V \to \ell^+ \ell^-$ ($V \to \pi \ell^+ \ell^-$) are significantly increased
(decreased). As a result, the two baselines present in the theory (see
Fig.~\ref{fig:ApBF}) become comparable in length, and viable parameter space is
reopened for $\epsilon \sim 10^{-3}$ and $\mAp \sim \text{few} \times 100
\text{ MeV}$, as shown in the bottom-left panel of  Fig.~\ref{fig:EpsAndFpi}.
However, larger values of $m_\pi / f_\pi$ also favor heavier thermal DM, shifting the
semi-annihilation and WIMP-like freeze-out regions to $m_\pi \gtrsim
\text{GeV}$. For $m_\pi / f_\pi = 4 \pi$ and $m_\pi \lesssim \text{GeV}$, freeze-out is instead controlled
by kinetic decoupling, as in Refs.~\cite{Kuflik:2017iqs,Kuflik:2015isi}. A
detailed discussion of such cosmology will appear in forthcoming work.
Large values of $m_\pi / f_\pi$ lead to a decrease in the branching fraction of $\Ap\rightarrow \pi\pi$, limiting the reach of experiments looking 
for missing energy signals, like BaBar, Belle-II, and LDMX. For $m_\pi/f_\pi=4\pi$, 
the signals at these experiments arise primarily from long-lived vector mesons decaying outside of 
the detectors. In the bottom-left panel of Fig.~\ref{fig:EpsAndFpi}, vector mesons often
decay visibly within the detectors at
Belle-II and LDMX for $\epsilon\sim 10^{-3}$.
As a result, the \textit{effective} invisible $\Ap$ branching fraction and the corresponding sensitivities of searches at BaBar, Belle-II, and LDMX are reduced.

In Fig.~\ref{fig:EpsAndFpi}, several parameters were held fixed. For the mass ratios considered here, variations in $\alpha_D$ do not qualitatively change our results. For instance, larger values of $\alpha_D$ lead to an overall decrease 
in the decay length of the vector mesons, which can be compensated by a decrease in the kinetic mixing, $\epsilon$. As a result, the
sensitivity contours of the fixed-target and beam dump experiments are correspondingly shifted down in the $m_\Ap-\epsilon$ plane, along with an overall suppression in the mass reach due to the suppressed production cross-section.
Different choices for the mass ratios 
can have more significant effects. For example, if $m_\Ap > 2 \, m_V$, then the new decay 
channel, $\Ap\rightarrow V^+ V^- \to \pi^+ \pi^-+ 4 \ell$, becomes kinematically accessible. 
Furthermore, the visible signals we have considered are suppressed if $\mAp \lesssim m_V + m_\pi$, 
although HS particle production is still possible through an off-shell dark photon.
In this case, direct detection, LDMX, and other missing momentum/energy experiments provide the most powerful probes 
of these models.

\section{Summary and Conclusions\label{sec:conclusion}}

Strongly interacting hidden sectors offer a compelling possibility for
accounting for the cosmological abundance of DM. This class of models requires
the existence of a coupling between the hidden sector (HS) and the SM to be cosmologically
viable. GeV-scale dark photons are ideal in this regard since kinetic mixing
with SM hypercharge facilitates equilibration between the
two sectors. Cosmology motivates scenarios in which the lightest spin-1 resonances of the HS are parametrically close in mass
to the DM pions. In this case, visible decays of these vector mesons
play an important role in the cosmological history of DM freeze-out. In this work, we
showed that $2 \to 1$ processes ($\pi \pi \to \pi V$ followed by $V \to
\text{SM}$) typically dominate over SIMP-like $3 \to 2$ annihilations and
greatly expand the allowed ranges of masses and couplings for thermal DM.

The presence of visibly-decaying vector mesons also opens a new avenue 
for experimental searches. The experimentally viable range of HS-SM couplings spans several orders of
magnitude, $\epsilon \sim 10^{-7} - 10^{-3}$, corresponding to a wide range of vector meson lifetimes. Thus, large regions of well-motivated
parameter space can be probed by searches for production and displaced decays of HS vector mesons. 
We showed that existing data from past beam dump and collider experiments already exclude some of these models. 
However, most of the cosmologically-motivated parameter space remains untested. 
Existing experiments like HPS will be able to cover new ground in this regard.

The wide range of viable vector meson lifetimes requires a multi-experiment approach to 
comprehensively cover the remaining low-mass parameter space. The next generation of fixed-target experiments in combination with long-lived particle searches 
at colliders will have unprecedented discovery reach for the vector mesons as well as the DM itself. 
For $\mAp \lesssim \text{few} \times \order{100} \text{ MeV}$, 
the proposed  LDMX experiment 
will be able to search for invisible dark photon decays of the form $\Ap \rightarrow \pi\pi$, as well as the displaced vertex signals discussed throughout this work.
For $\mAp \gtrsim \text{ GeV}$, Belle-II and an upgraded version of the existing SeaQuest spectrometer at FNAL will provide the most stringent tests. Above the muon threshold, similar searches at $B$-factories and the LHC can probe dark photons significantly heavier than a GeV.

In this paper, we have focused on a particular realization of a strongly interacting HS that mirrors SM QCD 
with three colors and three light flavors. However, the cosmology and the corresponding experimental signals 
are applicable to a much broader range of models with different gauge groups and flavor structures. This motivates ongoing and future experiments to test this interesting class of DM scenarios.

\section*{Acknowledgements}

We thank Jack Gunion, Yonit Hochberg, Anson Hook, Philip Ilten, Eder Izaguirre, John Jaros, Gordan Krnjaic, Eric Kuflik, Markus Luty, 
Gustavo Marques-Tavares, Michael Peskin,
Brian Shuve, and Yiming Zhong for useful conversations. We are especially
grateful to Takashi Maruyama and Bradley Yale for input regarding HPS
acceptances. AB, NB, PS, and NT are supported by
the U.S. Department of Energy under Contract No. DE-AC02-76SF00515. SG acknowledges support from the University of Cincinnati. SG is supported by a National Science Foundation CAREER Grant No. PHY-1654502.

\onecolumngrid
\begin{appendix}

\section{Model Details\label{app:model}}

In this Appendix, we briefly review the building blocks that are necessary for specifying interactions in the HS. Throughout this work, we have investigated the dynamics of a confining $SU(N_c)$ gauge theory, as discussed in Sec.~\ref{sec:model}. We focus on $N_c = 3$ colors and $N_f = 3$ flavors of light HS quarks in the fundamental representation of $SU(N_c)$, in analogy to SM QCD. The spontaneous breaking of the global chiral symmetry down to the diagonal subgroup proceeds as in the SM, i.e., $SU(N_f)_L \times SU(N_f)_R \to SU(N_f)_V$. Below the confinement scale, the $N_f^2 - 1$ broken generators correspond to the presence of pseudo-Nambu-Goldstone pions, $\pi^a$, which are the relevant low-energy degrees of freedom in the HS.

The interactions of the HS pions are described by the chiral Lagrangian, analogous to that of the SM. We write the exponentiated pion fields as
\be
U = \text{exp}\left( 2 i \pi^a T^a / f_\pi \right)
~,
\ee
where $T^a$ are the broken generators of $SU(N_f)$ and $f_\pi$ is the associated pion decay constant of the HS. We work in the convention where these generators are normalized to $\tr \left( T^a T^b \right) = \frac{1}{2} \delta^{ab}$. Since the number of colors and light flavors in the HS is identical to that of the light quark sector of the SM, we adopt the usual SM names for the HS mesons. We write the pion matrix as
\be
\Pi \equiv \pi^a T^a =  \frac{1}{\sqrt{2}} \begin{pmatrix} \frac{ \pi^0}{\sqrt{2}}  + \frac{\eta}{\sqrt{6}}  & \pi^+ & K^+ \\ \pi^- & -\frac{\pi^0}{\sqrt{2}}   + \frac{\eta}{\sqrt{6}}  & K^0 \\ K^- & \overline{K^0} & - \sqrt{\frac{2}{3}}  \, \eta \end{pmatrix}
.
\ee
The superscripts indicate charge under an additional $U(1)_D$ gauge symmetry as discussed below.
The kinetic and mass terms are
\be
\label{eq:chiralLag1}
\mathcal{L} \supset \frac{f_\pi^2}{4} \, \tr \left( \partial_\mu U \, \partial^\mu U^\dagger \right) + \frac{f_\pi^2 \, B_0}{2} \left(\tr \, M_q U^\dagger + \text{h.c.} \right)
,
\ee
where $M_q$ is the HS quark mass matrix and $B_0$ is a dimensionful scale. In the limit that $M_q \propto \mathbb{1}$, the mass of the degenerate pions is $m_\pi^2 = 2 \, B_0 \, M_q$ to leading order in chiral perturbation theory. 

We follow the phenomenological approach of Ref.~\cite{Klingl:1996by} in order to incorporate the spin-1 vector mesons into the low-energy theory. 
Adopting the SM naming convention, we parametrize the vector meson fields as
\be
V_\mu \equiv V_\mu^0 T^0 + V_\mu^a T^a = \frac{1}{\sqrt{2}} \begin{pmatrix} \frac{\rho^0}{\sqrt{2}} + \frac{\omega}{\sqrt{2}} &   \rho_\mu^+ &   K_\mu^{* +} \\  \rho_\mu^- & - \, \frac{\rho^0}{\sqrt{2}} + \frac{\omega}{\sqrt{2}} & K_\mu^{* 0}  \\  K_\mu^{* -} &  \overline{K_\mu^{* 0}}  &  \phi \end{pmatrix}
,
\ee
where in the sum we have included the diagonal generator, $T^0 = 1 / \sqrt{6}$, corresponding to the $U(1)$ subgroup of the $U(N_f)$ flavor symmetry. The kinetic terms for the vector mesons are given by
\be
\mathcal{L} \supset - \, \frac{1}{2} ~ \tr ( V_{\mu \nu} V^{\mu \nu} )
~,
\ee
where the field strength is $V_{\mu \nu} = \partial_\mu V_\nu - \partial_\nu V_\mu - i g \left[ V_\mu , V_\nu \right]$. We assume that the vector mesons share a common mass, $m_V$, unless stated otherwise. As mentioned in Sec.~\ref{sec:model}, the KSRF relation implies that the vector meson coupling, $g$, can be estimated as $g \simeq m_V/(\sqrt{2} \, f_\pi)$. We have utilized this parametrization throughout. This coupling also directly enters into the $V \pi \pi$ interaction by promoting the derivative of the pion field to the covariant form 
\be
\partial_\mu \Pi \to D _\mu \Pi \supset \partial_\mu \Pi + i g~  [ \Pi , V_\mu]
~.
\ee

Interactions of the dark photon, $\Ap$, with the HS mesons are incorporated by gauging a $U(1)_D$ subgroup of the unbroken diagonal flavor symmetry. We add to the covariant pion derivative, 
\be
\label{eq:pionDerivAp}
D _\mu \Pi \supset i e_D [ \Pi, Q] A^\prime_\mu
~,
\ee
where $e_D$ is the $U(1)_D$ gauge coupling and $Q$ is the $U(1)_D$ charge matrix of the HS quarks, as specified in Eq.~\eqref{eq:Q}. The vector mesons acquire a coupling to the $\Ap$ through kinetic mixing~\cite{Klingl:1996by,Hochberg:2015vrg}
\be
\label{eq:vectorApKinMix}
\mathcal{L} \supset - \, \frac{e_D}{g} ~ A_{\mu \nu}^\prime ~ \tr \, Q \, V_{\mu \nu} 
~.
\ee
Alternatively, interactions between the $\Ap$ and HS mesons can be implemented in the Vector Meson Dominance limit of the Hidden Local Symmetry framework~\cite{Harada:2003jx}. In this case, the $\Ap$ only couples to $\pi^\pm$ and $V^\pm$ pairs through intermediate neutral vector mesons, as in Eq.~\eqref{eq:vectorApKinMix}.

\section{Decays\label{app:dec}}

In this Appendix, we provide expressions for the decays of the $\Ap$ and HS vector mesons. This is especially relevant for the experimental signals outlined in Sec.~\ref{sec:signals}. For the sake of brevity, we introduce the notation
\be
r_i \equiv m_i / \mAp
~,
\ee
where $i = \ell, \pi, V$ denotes a SM lepton, HS pion, or HS vector meson, respectively. 

The $\Ap$ can decay to SM fermions through kinetic mixing with SM hypercharge. In the limit that $\mAp \ll m_Z$, the width for the decay of an $\Ap$ into a pair of SM leptons is approximated by
\be
\Gamma (\Ap \to \ell^+ \ell^-) \simeq \frac{ \alpha_\text{em} \, \epsilon^2}{3} \, \left( 1 - 4 \, r_\ell^2 \right)^{1/2} ~ \left( 1 + 2 \, r_\ell^2 \right) \mAp 
~.
\ee
We incorporate the decays of the dark photon to SM hadrons through the data-driven parameter $R(\sqrt{s}) \equiv \sigma (e^+ e^- \to \text{hadrons}) / \sigma (e^+ e^- \to \mu^+ \mu^-)$, as in Ref.~\cite{Patrignani:2016xqp}. In particular, 
\be
\label{eq:ApToHadrons}
\Gamma (\Ap \to \text{SM hadrons}) \simeq R(\sqrt{s} = \mAp) ~ \Gamma (\Ap \to \mu^+ \mu^-)
~.
\ee

If $\mAp \gtrsim 2 \, m_\pi$, $\Ap$ can decay to pairs of HS pions that are directly charged under $U(1)_D$. From Eq.~\eqref{eq:pionDerivAp}, the corresponding width is
\be
\Gamma (\Ap \to \pi \pi) = \frac{2 \, \alpha_D}{3} ~ \frac{(1 - 4 r_\pi^2)^{3/2}}{(1-r_V^{-2})^2} ~ \mAp
~,
\ee
where $\alpha_D \equiv e_D^2 / 4 \pi$. The interactions between the $\Ap$ and pairs of vector mesons follow in a similar manner, leading to
\be
\Gamma (\Ap \to V V) = \frac{\alpha_D}{6} ~ \frac{(1 - 4 r_V^2)^{1/2} (1 + 16 r_V^2 - 68 r_V^4 - 48 r_V^6)}{(1 - r_V^2)^2} ~ \mAp
~.
\ee
For $\mAp \gtrsim m_\pi + m_V$, $\Ap$ also decays to $\pi V$ final states. The calculation for such processes is analogous to determining $\pi^0_\text{SM} \to \gamma \gamma$ in the SM and follows from the standard anomalous chiral current algebra. We find
\begin{align}
\Gamma (\Ap \to \eta ~ \rho^0 ) &= \frac{\alpha_D \, r_V^2}{256 \pi^4} ~ \left( \frac{m_\pi / f_\pi}{r_\pi} \right)^4 ~ \Big[ 1 - 2 (r_\pi^2 + r_V^2) + (r_\pi^2 - r_V^2)^2 \Big]^{3/2} ~ \mAp
\nl
\Gamma (\Ap \to \eta ~ \phi) &= 2 \times \Gamma (\Ap \to \eta ~ \rho^0 )
\nl
\Gamma (\Ap \to \pi^0 ~ \omega) &= 3 \times \Gamma (\Ap \to \eta ~ \rho^0 )
\nl
\Gamma (\Ap \to K^0 ~ \overline{K^{*0}} ~,~ \overline{K^0} ~ K^{*0}) &= 6 \times \Gamma (\Ap \to \eta ~ \rho^0 )
\nl
\Gamma (\Ap \to \pi^\pm ~ \rho^\mp) &= 6 \times \Gamma (\Ap \to \eta ~ \rho^0 )
\nl
\Gamma (\Ap \to K^\pm ~ K^{*\mp}) &=6 \times \Gamma (\Ap \to \eta ~ \rho^0 )
~.
\end{align}

If $m_V \lesssim 2 \, m_\pi$, then vector mesons can only decay to the SM. If $\tr \left[ Q \, T^a \right] \neq 0$ as well, this decay proceeds through $V^a$ kinetically mixing with $\Ap$, as in Eq.~\eqref{eq:vectorApKinMix}, and $V^a$ dominantly decays to pairs of SM fermions. In the limit that $m_V, \mAp \ll m_Z$, the corresponding decays to SM leptons are given by
\begin{align}
\label{eq:2bodyDecay}
\Gamma (\rho^0 \to \ell^+ \ell^-) &\simeq \frac{32 \pi \, \alpha_\text{em} \, \alpha_D \, \epsilon^2}{3} ~ \left( \frac{r_\pi}{m_\pi / f_\pi} \right)^2 \, (r_V^2 - 4 r_\ell^2)^{1/2} \, (r_V^2+2r_\ell^2) \, (1- r_V^2)^{-2} ~ \mAp
\nl
\Gamma (\phi \to \ell^+ \ell^-) &\simeq \frac{1}{2} \times \Gamma (\rho^0 \to \ell^+ \ell^-) 
\nl
\Gamma (\omega \to \ell^+ \ell^-) &\simeq 0
~.
\end{align}
Similar to Eq.~\eqref{eq:ApToHadrons}, we can also express the decay of vector mesons into SM hadrons through
\be
\Gamma (V \to \text{SM hadrons}) = R(\sqrt{s} = m_V) ~ \Gamma (V \to \mu^+ \mu^-)
~.
\ee

On the other hand, if $m_V \lesssim 2 \, m_\pi$ but $\tr \left[ Q \, T^a \right] = 0$, then $V^a$ decays through a three-body process involving an off-shell $\Ap$, i.e., $V^a \to \pi^b A^{\prime *} \to \pi^b \ell^+ \ell^-$, as discussed in Sec.~\ref{sec:model}. In the limit that $m_{\ell} \ll m_V , \mAp \ll m_Z$, the non-zero widths are approximated as
\begin{align}
\Gamma (\rho^0 \to \eta \ell^+ \ell^- ) &\simeq \frac{\alpha_\text{em} \, \alpha_D  \, \epsilon^2 \, (m_\pi / f_\pi)^4 \, \mAp }{4608 \pi^5 \, r_\pi^4 \, r_V \, \Big[ r_\pi^4 - 2 \, r_\pi^2 (1 + r_V^2 ) + (1 - r_V^2 )^2 \Big]^{1/2}}
\Bigg\{ \bigg[3 \left(r_V^2-r_\pi^2\right)^4-21 \left(r_\pi^2+r_V^2\right) \left(r_V^2-r_\pi^2\right)^2
\nl
& -39 \left(r_\pi^2+r_V^2\right) +15 \left(3 r_\pi^4+3 r_V^4+2 r_\pi^2 r_V^2\right)+12\bigg] 
\nl
& \times \log{\frac{\left( 1 + \frac{-r_\pi^2+r_V^2+1}{\big[\left(r_\pi^2-1\right)^2+r_V^4-2 \left(r_\pi^2+1\right) r_V^2 \big]^{1/2}}\right) \left(1-\frac{r_\pi^2-r_V^2+1}{\big[\left(r_\pi^2-1\right)^2+r_V^4-2 \left(r_\pi^2+1\right) r_V^2\big]^{1/2}}\right)}{\left(1-\frac{-r_\pi^2+r_V^2+1}{\big[\left(r_\pi^2-1\right)^2+r_V^4-2 \left(r_\pi^2+1\right) r_V^2\big]^{1/2}}\right) \left(1+\frac{r_\pi^2-r_V^2+1}{\big[\left(r_\pi^2-1\right)^2+r_V^4-2 \left(r_\pi^2+1\right) r_V^2\big]^{1/2}}\right)}}
\nl
&+\big[r_\pi^4-2 r_\pi^2 \left(r_V^2+1\right)+\left(r_V^2-1\right)^2\big]^{1/2} \bigg[17 \left(r_\pi^6- r_V^6 \right)+42 \left(r_V^4-r_\pi^4\right)+39 \left(r_\pi^2 r_V^4-r_\pi^4 r_V^2\right)+24 \left(r_\pi^2-r_V^2\right)
\nl
&+\Big(36 \left(r_\pi^4+r_V^4\right)-54 \left(r_\pi^2+r_V^2\right)-6 \left(r_\pi^4+r_V^4-4 r_\pi^2 r_V^2\right) \left(r_\pi^2+r_V^2\right)+24\Big) \log{\frac{r_V}{r_\pi}}\bigg] \Bigg \}
\end{align}
and
\begin{align}
\Gamma (\phi \to \eta \ell^+ \ell^- ) &\simeq 2 \times \Gamma (\rho^0 \to \eta \ell^+ \ell^- )
\nl
\Gamma (\omega \to \pi^0 \ell^+ \ell^- ) &\simeq 3 \times \Gamma (\rho^0 \to \eta \ell^+ \ell^- )
\nl
\Gamma (K^{* 0} \to K^0 \ell^+ \ell^-) &\simeq 3 \times \Gamma (\rho^0 \to \eta \ell^+ \ell^- )
\nl
\Gamma (\overline{K^{* 0}} \to \overline{K^0}  \ell^+ \ell^-) &\simeq 3 \times \Gamma (\rho^0 \to \eta \ell^+ \ell^- )
\nl
\Gamma (\rho^\pm \to \pi^\pm \ell^+ \ell^-) &\simeq 3 \times \Gamma (\rho^0 \to \eta \ell^+ \ell^- )
\nl
\Gamma (K^{* \pm} \to K^\pm \ell^+ \ell^-) &\simeq 3 \times \Gamma (\rho^0 \to \eta \ell^+ \ell^- )
~.
\end{align}

\section{Boltzmann Equations\label{app:cosmo}}

Once the temperature drops below the DM pion mass, $T \lesssim m_\pi$, various processes maintain chemical equilibrium between the HS and SM. These include pion annihilation into SM fermions, $f$, through an intermediate dark photon ($\pi \pi \to A^{\prime *} \to f \bar{f}$), $3 \to 2$ pion scattering ($\pi \pi \pi \to \pi \pi$), and semi-annihilation into vector mesons ($\pi \pi \to \pi V$), which later decay to the SM ($V \to f \bar{f}$). Since $m_V \gtrsim m_\pi$ implies that $\pi \pi \to \pi V$ is kinematically suppressed at small temperatures, we refer to this latter process as ``forbidden semi-annihilation." 

The Boltzmann equations governing the number density and temperature evolution of the HS can be written as
\begin{align}
\label{eq:boltzmann}
\dot{n}_\pi + 3 H n_\pi &= %2 ~ \langle \Gamma (V \to \pi \pi) \rangle (T_\text{HS}) ~~ \Big[ \, n_V - (n_\pi / n_\pi^\text{eq})^2 ~ n_V^\text{eq} (T_\text{HS}) \, \Big] 
%\nl
%& +
n_\pi ~ \langle \sigma v (\pi V \to \pi \pi) \rangle (T_\text{HS}) ~~ \Big[ \, n_V - (n_\pi / n_\pi^\text{eq}) ~~ n_V^\text{eq} (T_\text{HS}) \, \Big]
\nl
& - n_\pi^2 ~ \langle \sigma v^2 (\pi \pi \pi \to \pi \pi )\rangle (T_\text{HS}) ~~ \Big[ \, n_\pi - n_\pi^\text{eq} (T_\text{HS}) \, \Big] 
\nl
& - \Big[ \,  \langle \sigma v (\pi \pi \to \text{SM} ) \rangle (T_\text{HS}) ~~ n_\pi^2 ~ - ~ \langle \sigma v (\pi \pi \to \text{SM} ) \rangle (T_\text{SM}) ~~ (n_\pi^{\text{eq}})^2 (T_\text{SM} )  \, \Big],
\nl
\nl
\dot{n}_V + 3 H n_V &= - \Big[ \, \langle \Gamma (V \to \text{SM}) \rangle (T_\text{HS}) ~~ n_V - \langle \Gamma (V \to \text{SM}) \rangle (T_\text{SM}) ~~ n_V^\text{eq} (T_\text{SM})  \, \Big] 
\nl
& - n_\pi ~ \langle \sigma v (\pi V \to \pi \pi) \rangle (T_\text{HS}) ~~ \Big[ \, n_V - (n_\pi / n_\pi^\text{eq}) ~~ n_V^\text{eq} (T_\text{HS}) \, \Big],
\nl
\nl
\dot{\rho}_{\pi + V} + 3 H \left( \rho_{\pi + V} + P_{\pi + V} \right) &= - m_V ~ \Gamma (V \to \text{SM}) ~ \Big[ \, n_V - n_V^\text{eq} (T_\text{SM}) \, \Big] 
\nl
& - \langle \sigma v (\pi \, \text{SM} \to \pi \, \text{SM}) ~ \Delta E \rangle ~~ n_\text{SM}^\text{eq}(T_\text{SM}) ~~ n_\pi
~.
\end{align}
Above, brackets denote thermal averaging, and we have been explicit at which temperature various quantities are to be evaluated. 
The equilibrium distributions are approximated assuming Maxwell-Boltzmann statistics, 
\be
n^\text{eq} \simeq \frac{g \, m^2 \, T}{2 \, \pi^2} ~ K_2 (m / T)
~,~
\rho^\text{eq} \simeq \frac{g \, m^2 \, T}{2 \, \pi^2} ~ \left( m \, K_1 (m / T) + 3 \, T \, K_2 (m / T) \right)
~,~
P^\text{eq} \simeq T \, n^\text{eq}
~,
\ee
where $K_{1,2}$ are modified Bessel functions of the second kind, and $n_{\pi (V)}^\text{eq}$ is evaluated with $g=8$ (9) flavor degrees of freedom. We numerically solve Eq.~\eqref{eq:boltzmann} for $n_\pi (T_\text{SM})$, $n_V (T_\text{SM})$, and $T_\text{HS} (T_\text{SM})$ after substituting
\be
\frac{d}{dt} \simeq - H ~ T_\text{SM} ~ \frac{d}{d \, T_\text{SM}}
~,~
\rho \simeq \frac{n}{n^\text{eq}} ~ \rho^\text{eq}
~,~ 
P \simeq \frac{n}{n^\text{eq} } ~ P^\text{eq}  \simeq T \, n
~.
\ee

We now provide expressions for the effective decay and annihilation rates that enter into Eq.~\eqref{eq:boltzmann}. In the limit that $m_V \gg m_\ell$, the thermally-averaged decay rate into SM leptons is given by
\be
\langle \Gamma (V \to \ell^+ \ell^-) \rangle \simeq \frac{K_1(m_V/T)}{K_2(m_V / T)} ~ \frac{16 \, \pi  \, \alpha_D \, \alpha_\text{em} \, \epsilon^2 \, m_\pi^2 \, m_V^3}{9 \, (m_\pi / f_\pi)^2 \left(\mAp^2-m_V^2\right)^2}
~.
\ee
We also include decays to hadrons in Eq.~\eqref{eq:boltzmann}, as described in Appendix~\ref{app:dec}.
In the non-relativistic limit, the thermally-averaged cross-section for DM annihilation into SM leptons is approximately,
\be
\langle \sigma v (\pi \pi \to \ell^+ \ell^-)\rangle \simeq 4 \pi  \, \alpha_D \, \alpha_\text{em} \, \epsilon^2 ~ \frac{m_\pi \, m_V^4 \, T}{\left(4 m_\pi^2 - \mAp^2\right)^2 \left(4 m_\pi^2 - m_V^2 \right)^2}
 ~,
\ee
where we have again taken $m_\ell \simeq 0$. In our analysis, we numerically evaluate the general form of $\langle \sigma v (\pi \pi \to \ell^+ \ell^-)\rangle$, following the procedure outlined in Refs.~\cite{Griest:1990kh,Edsjo:1997bg}. We also incorporate direct annihilations to SM hadrons, as discussed in Ref.~\cite{Izaguirre:2015zva}. In evaluating $\langle \sigma v^2 (\pi \pi \pi \to \pi \pi)\rangle$ and $\langle \sigma v (\pi \, \text{SM} \to \pi \, \text{SM}) ~ \Delta E \rangle$, we utilize the results in Ref.~\cite{Hochberg:2014kqa} and Ref.~\cite{Kuflik:2017iqs}, respectively. We follow Refs.~\cite{Kaymakcalan:1983qq,Gondolo:1990dk} for the calculation of the semi-annihilation process. In the non-relativistic limit, we find 
\begin{align}
 \langle \sigma v (\pi V \to \pi \pi)\rangle &\simeq \frac{5 \, (m_\pi / f_\pi)^8 \, m_V^2 \, \big[(m_V - m_\pi) (3 \, m_\pi + m_V)\big]^{3/2} \, T}{1179648 \, \pi ^5 \, m_\pi^{10}  \, (m_\pi + 2 \, m_V)^2 \left(m_V^2 -m_\pi^2 + m_\pi \, m_V \right)^2}
\nl  
& \times \Big( m_\pi^8 + 24 \, m_V^8 - 64 \, m_\pi \, m_V^7 + 148 \, m_\pi^2 \, m_V^6 + 428 \,m_\pi^3 \, m_V^5 
\nl
&\quad ~~ + 205 \, m_\pi^4 \, m_V^4 - 58 \, m_\pi^5 \, m_V^3 - 29 \, m_\pi^6 \, m_V^2 + 2 \, m_\pi^7 \, m_V \Big)
~.
\end{align}
If the vector mesons that do \textit{not} undergo two-body decays to SM particles are decoupled from the low-energy theory, as discussed in Sec.~\ref{sec:signals}, then various quantities in Eq.~\eqref{eq:boltzmann} are rescaled, 
\be
n_V^\text{eq} \to \frac{2}{9} ~ n_V^\text{eq}
~,~
\langle \sigma v (\pi V \to \pi \pi)\rangle \to \frac{9}{8} ~ \langle \sigma v (\pi V \to \pi \pi)\rangle
~,~
\langle \Gamma (V \to \text{SM}) \rangle \to \frac{9}{2} ~ \langle \Gamma (V \to \text{SM}) \rangle
~.
\ee

\end{appendix}

\twocolumngrid
\bibliography{SIMP}

\end{document}